\documentclass[]{WileyASNA-v1}
\usepackage{bm}
\usepackage{amsmath}
\usepackage{lipsum}  

\usepackage{color}

\setcitestyle{aysep={},yysep={;},maxcitenames=2}

 \renewcommand{\thefigure}{\arabic{figure}}
 \let\oldcaption=\caption
 \renewcommand{\caption}[1]{\oldcaption{#1}}

\newcommand*\arcsec{\ensuremath{^{\prime\prime}}}
\DeclareRobustCommand{\ion}[2]{\textup{#1\,\textsc{\lowercase{#2}}}}

\DeclareMathOperator\erf{erf}

\articletype{Research Article}%

\received{...}
\revised{...}
\accepted{...}

\newcommand{\mytitle}{Revisiting the Emission Line Source Detection Problem in Integral
  Field Spectroscopic Data}

\begin{document}

\title{\mytitle}

\author{Edmund Christian Herenz}

\authormark{E.~C.~HERENZ }

\address{\orgname{Fellow with the European Southern Observatory},
  \orgaddress{Av. Alonso de C\'ordova 3107, 763 0355 Vitacura,
    \state{Santiago}, \country{Chile}}}

\presentaddress{Leiden Observatory, Leiden University, Niels Bohrweg 2, NL 2333
  CA Leiden, The Netherlands}


\corres{\email{christian@herenz.cc}}

\abstract{We present a 3-dimensional matched filtering approach for the blind
  search of faint emission-line sources in integral-field spectroscopic
  datasets.  The filter is designed to account for the spectrally rapidly
  varying background noise due to the telluric air glow spectrum.  A software
  implementation of this matched filtering search is implemented in an updated
  version of the Line Source Detection Cataloguing tool (LSDCat2.0).  Using
  public data from the MUSE-Wide survey we show how the new filter design
  provides higher detection significances for faint emission line sources buried
  in between atmospheric [OH]-bands at $\lambda \gtrsim 7000$\,\AA{}.  We also
  show how, for a given source parameterisation, the selection function of the
  improved algorithm can be derived analytically from the variances of the data.
  We verify this analytic solution against source insertion and recovery
  experiments in the recently released dataset of the MUSE eXtreme Deep Field
  (MXDF).  We then illustrate how the selection function has to be rescaled for
  3D emission line source profiles that are not fully congruent with the
  template.  This procedure alleviates the construction of realistic selection
  functions by removing the need for computationally cumbersome source insertion
  and recovery experiments. }



\jnlcitation{\cname{%
\author{E.~C.~Herenz}} (\cyear{2022}), 
\ctitle{\mytitle}, \cvol{2022;XXX:X--Y}.}

\maketitle



\defcitealias{Herenz2017}{HW17}

\section{Introduction}
\label{sec:intro}

Modern wide-field integral-field spectrographs \citep[IFS;][]{Bacon2017a} on
large ground based telescopes offer unprecedented capabilities for spectroscopic
surveys of faint emission line galaxies.  Currently these instruments are
represented by the Multi-Unit Spectroscopic Explorer on ESO's Very Large
Telescope ``Yepun'' at Cerro Paranal \citep[MUSE;][]{Bacon2014} and the Keck
Cosmic Web Imager on Keck II at Manua Kea \citep{Morrissey2018}.  The continuous
spatial and spectral coverage provided by wide-field IFS avert the classical
spectroscopic survey paradigm of target pre-selection for follow-up
spectroscopy, as these observations provide literally ``spectroscopy of
everything'' within the observed fields.  This concept is also utilised in the
Hobby Eberly Telescope Dark Energy Experiment
\citep[HETDEX;][]{Gebhardt2021,Hill2021}, a dedicated IFS survey facility for
Lyman $\alpha$ (Ly$\alpha$\,$\lambda1216$) emitting galaxies (LAEs;
\citealt{Ouchi2019}), that aims at constraining the Dark Energy equation of
state.  Given their pivotal importance in addressing fundamental astrophysical
and cosmological questions the next generation of wide-field IFS are already
envisioned \citep[e.g., as part of the ESO Spectroscopic
Facility;][]{Pasquini2018} or in early stages of planning \citep[BlueMUSE for
ESO/VLT;][]{Richard2019}.

Efficient detection and cataloguing of sources in IFS datasets is obviously a
crucial task for fully exploiting the survey capabilities of these instruments.
This analysis step follows in sequence to the data reduction of the
observational raw data.  The data reduction pipelines processes the instrumental
detector data of the on-sky exposures into science-ready data-structures (see,
e.g., \citealt{Weilbacher2020} for the MUSE data reduction pipeline and
\citealt{Turner2010} for general aspects of IFS data reduction).  The reduced
high-level data-products are three dimensional (3D) arrays, with two axes
mapping the spatial domain (on-sky position / Right Ascension and Declination)
and the third axis mapping the spectral domain (wavelength).  One element of
this 3D array is called a voxel (short form of volume pixel), and the collection
of all voxels at the same spatial coordinate are called spaxel (spectral pixel).
Each voxel stores a scalar quantity related to the flux density at this
wavelength and position on the sky.  The jargon for such IFS data-products is
``datacubes'' in literature \citep[e.g., review by][]{Allington-Smith2006}, but
since generally the length of the spatial and spectral axes differ we will here
use the formally correct term ``data cuboid''.

In the last decade several works were concerned with the problem of detecting
and cataloguing faint emission line sources in data cuboids from deep MUSE
observations \citep[e.g.,][]{Bourguignon2012}.  This happened because already
first science case that motivated the construction of MUSE framed it as a
discovery machine for the faintest line emitting objects. \citep{Bacon2002}; a
dream that has now become reality
\citep[e.g.,][]{Maseda2018,Bacon2021,SanchezAlmeida2022}.  As of this writing
the following software implementations for detecting and cataloguing emission
lines in wide-field IFS data cuboids are freely\footnote{Another software is
  \texttt{CubeEx}, but this programme is only available from the author on
  request \citep[see the the ``code availability'' remark in][]{Ginolfi2022}.}
available: ``Source Emission Line FInder'' \citep[SELFI;][]{Meillier2016},
ORIGIN \citep{Mary2020}, MUSELET \citep[a wrapper of the popular imaging source
detection software SExtractor and part of the MUSE Python Data Analysis
Framework;][]{Piqueras2017}, and the ``Line Source Detection and Cataloguing
Tool'' \mbox{LSDCat} \citep[][hereafter
HW17]{Herenz2017}.

LSDCat distinguishes itself from ORIGIN or SELFI by a rather simplistic
detection and cataloguing approach.  It is based around a straightforward
two-step process of 3D linear (matched) filtering and thresholding; an optional
third step can be invoked to parameterise the found emission line sources.  The
simplicity allows for fast processing of MUSE data cuboids, which comprise
typically data volumes of $\sim 3$\,GByte per pointing (flux data and
corresponding variances).  Fast processing is a requirement when working with a
large number of such data cuboids, but it allows also for efficient source
insertion and recovery experiments to empirically determine the selection
function for some particular type of emission line galaxies.  Robust selection
function determinations are a must when estimating and modelling fundamental
galaxy population statistics.  Worked out examples in this respect are the
spatial clustering measurement \citep{HerreroAlonso2021} as well as the
luminosity function determination \citep{Herenz2019} of Ly$\alpha$ emitting
galaxies at redshifts $3 \lesssim z \lesssim 6$ in the MUSE-Wide survey.

In the present article we present a significant improvement of the filtering
algorithm of \mbox{LSDCat}\footnote{LSDCat can be obtained from the Astrophysics
  Source Code Library: \url{https://ascl.net/1612.002}.}.  The improved method
accounts for high-frequency variations of the noise in spectral direction within
IFS data cuboids.  A high incidence of such high-frequency variations at
considerable strength occurs at $\lambda \gtrsim 7000$\,\AA{} in ground based
observations due to the atmospheric spectral lines of OH-molecules \citep[see
Sect.~4.3 in][]{Noll2012}.  Our revision of the original algorithm is motivated
primarily by the desire to improve the detectability of the faintest Ly$\alpha$
emitting galaxies at the highest redshifts in the deepest MUSE-datasets, i.e.,
the MUSE Hubble Ultra Deep Field
\citep[$t_\mathrm{exp} \gtrsim 30$\,h;][]{Bacon2017,Inami2017} and the MUSE
eXtreme Deep Field \citep[MXDF, $t_\mathrm{exp} =141$\,h;][]{Bacon2022}.  Here
it was found that robust detections with ORIGIN appear not as significantly
detected with LSDCat while others could even be missed at reasonable detection
thresholds (R.~Bacon, priv. comm.).  Another motivation is to provide a
deterministic detection algorithm, whose selection function is solely defined by
the noise properties of the data and the physical properties of the coveted
emission line sources (i.e., their line flux, as well as their spatial-, and
spectral profiles).

While this article accompanies the release of the improved version of the
\mbox{LSDCat} software (\mbox{LSDCat2.0}), its aim is also to provide a robust
general formal framework for the detection of emission line sources in
integral-field spectroscopic datasets via matched filtering.  The remainder of
this article is structured as follows: In Sect.~\ref{sec:2} we review some basic
properties of source detection by matched filtering that are of relevance for
the task at hand (Sect.~\ref{sec:mf}).  We then briefly review the original
algorithm in LSDCat (Sect.~\ref{sec:classic}) before explaining the
modifications in the improved version (Sect.~\ref{sec:modified}).  Next we
contrast the improved algorithm to the previous version and we verify the
implementation against the detected emission lines within the first public data
release of the MUSE-Wide survey (Sect.~\ref{sec:verify}).
Section~\ref{sec:selfun} then describes the construction of the deterministic
selection function of the detection metod.  First, we derive the selection
function for the idealised case where the spectral profiles of the surveyed
emission line sources are exactly known and where the spatial profiles are
indistinguishable from the instrumental point spread function
(Sect.~\ref{sec:ideal-select-funct}).  This idealised selection function is then
verified against an empirical source insertion and recovery experiment in the
recently released MUSE data cuboid of the MUSE eXtreme Deep Field
(Sect.~\ref{sec:verif-analyt-expr}).  Lastly, we present a rescaling method of
the idealised selection function to different source profiles
(Sect.~\ref{sec:rs}).  We close the paper with a summary and conclusion in
Sect.~\ref{sec:sc}.

\section{An Improved 3D Line Detection Algorithm}
\label{sec:2}

Source detection is, at its heart, a statistical test to rule out the hypothesis
$\mathcal{H}_0$ of no source being present somewhere in the observational
dataset under examination.  If $\mathcal{H}_0$ can be rejected with confidence,
we accept the reasonable alternative, $\mathcal{H}_1$, that the signature from
the sought astronomical source is present in our dataset.  \mbox{LSDCat}, like
virtually every astronomical source detection software, is based around linear
filtering of the data prior to deciding between $\mathcal{H}_1$ and
$\mathcal{H}_0$.

Arguably, the most suitable filter for the detection of isolated astronomical
emission line sources in a background noise dominated observational dataset is
the ``matched filter''; the name resonates that the response of the filter is
matched to the known or expected shape of the signal to be detected.  Matched
filters provide the optimal test statistic to decide between $\mathcal{H}_0$ and
$\mathcal{H}_1$, i.e., no other statistic provides a better decision criterion
\citep[see, e.g.,][for a textbook treament of matched filtering]{Das1991}.  The
applications of the matched filter are widespread in science and engineering,
e.g., detection of gravitational waves \citep{Jaranowski2012} or radar
\citep{Schwartz1975}, and in the following we describe how this concept is being
applied to the emission line source detection problem in IFS data.  A recent
review on advanced matched filtering concepts relevant for astronomy was
presented by \cite{Vio2021} and we follow in some parts the notation developed
in there.

\subsection{Matched Filtering for Emission Lines in IFS data cuboids}
\label{sec:mf}

Analysing first the one-dimensional discrete case with Gaussian noise allows us
to formalise the line detection problem as follows.  We consider a line profile
that is represented by a shape vector
\begin{equation}
  \label{eq:1}
  \bm{s} = (s_1, \dots, s_M)^T  \;\text{,}
\end{equation}
which is normalised, i.e.,  $ \sum_i s_i \equiv 1$.  This normalised shape,
multiplied by some amplitude $a > 0$, may or may not be present in a noisy data
vector
\begin{equation}
  \label{eq:2}
  \bm{f} = (f_1, f_2, \dots, f_i, \dots, f_{N-1}, f_N)^T
\end{equation}
at known position $i_0$.  The
noise of $\bm{f}$ is given by a Gaussian noise vector with zero mean,
\begin{equation}
  \label{eq:3}
  \bm{n}=(n_1, \dots, n_N)^T \;\text{,}
\end{equation}
that relates\footnote{For example, given a covariance matrix $\bm{\Sigma}$ a
  realisation the noise vector $\bm{n}$ follows from the decomposition
  $\bm{\Sigma} = \bm{C} \bm{C}^T$ and a vector of standard normal uncorrelated
  noise $\tilde{\bm{n}}$ via $\bm{n} = \bm{C} \tilde{\bm{n}}$.} to an
$N\times N$ covariance matrix $\bm{\Sigma}$.  In Eqs.~\eqref{eq:1} to
\eqref{eq:3} and below the superscript $T$ indicates the transpose operation.

Explicitly, we here think of $\bm{f}$ in Eq.~\eqref{eq:2} as an observed 1D
spectrum with $i = 1, \dots, N$ indexing the grid of spectral bins, while
$\bm{s}$ in Eq.~\eqref{eq:1} describes the known (or expected) shape of the
emission line to be detected.  The position $i_0$ refers to some well defined
feature in $\bm{s}$, e.g., the position of the line peak or the first moment of
the line profile.  For the problem to be well defined the line width of $\bm{s}$
is assumed to be significantly smaller compared to the length of $\bm{f}$, i.e.,
$M \ll N$.  Lastly, $\bm{n}$ in Eq.~\eqref{eq:3} describes the noise of the
spectrum that results from the detector (read out and dark current) and, for
ground based observations, shot noise from the telluric background\footnote{The
  natural sources of the telluric background are described in
  \cite{Noll2012}. Artificial components are due to light pollution
  \citep{Green2022} and, in case of laser assisted adaptive optics observations,
  due to laser-induced Raman scattered photons by molecules in the atmosphere
  \citep{Vogt2017,Vogt2019}.} \citep[e.g.,][]{Ferruit2010}.  The off-diagonal
terms in $\bm{\Sigma}$ that define the co-variance in $\bm{n}$ are caused by
resampling of the detector pixels, e.g., from multiple exposures and/or
rectifying procedures, onto the final data grid within the data reduction
pipeline.  We point out that we here strictly assume the that the emission line
source does not contribute to the noise background, that is we assume the shot
noise from the source is negligible in comparison to the other noise
contributions mentioned above.

Without loss of generality we now fix $M \bmod 2 = 1$ and assert that $i_0$
corresponds to the central pixel of the line signal in $\bm{f}$.  This allows us to
introduce a vector of length $N$, in which $\bm{s}$ is embedded and adequately
padded with zeros, i.e., 
\begin{equation}
  \label{eq:4}
  \bm{s}^{(i_0)} =
  \left (
    \bm{0}_{[i_0 - \kappa - 1]}^T \, ,
    \,
    \bm{s}^T \, ,
    \,
    \bm{0}_{[N - i_0 - \kappa]}^T
  \right )^T  \;\text{,}  
\end{equation}
where $\bm{0}_{[x]}$ denotes a zero vector of length $x$ and where we introduced
the shorthand $\kappa = \frac{M-1}{2}$ for ease of notation.  So called ``edge
effects'' occur for $i_0 \leq \kappa - 1$ and $i_0 \geq N - \kappa $, and for
such lines the detection problem is not well defined.

In order to argue for the presence or absence of a source signal at $i_0$ we now
require a statistic to decide between $\mathcal{H}_0$ and $\mathcal{H}_1$, formally
\begin{align}
  \label{eq:4mod}
  \begin{split}
    \mathcal{H}_0 &\Leftrightarrow \bm{f} = \bm{n}    \\ \text{or} \quad
    \mathcal{H}_1 &\Leftrightarrow \bm{f} = a \cdot \bm{s}^{(i_0)} + \bm{n} \;\text{.}
  \end{split}
\end{align}
It is proven with mathematical rigour that the optimal
decision statistic for rejecting $\mathcal{H}_0$ based on the alternative
$\mathcal{H}_1$ in Eq.~\eqref{eq:4mod} is given by the scalar product
\begin{equation}
  \label{eq:5}
  \mathcal{T}^{(i_0)} = \left (\bm{T}^{(i_0)} \right )^T \bm{f} \;\text{,}
\end{equation}
with the vector
\begin{equation}
  \label{eq:6}
  \bm{T}^{(i_0)} = \alpha_{(i_0)} \, \bm{\Sigma}^{-1} \bm{s}^{(i_0)} 
\end{equation}
being the \textit{matched filter} \citep[e.g.,][]{Vio2021}.  $\bm{\Sigma}^{-1}$
in Eq.~\eqref{eq:6} denotes the inverse of the co-variance matrix $\bm{\Sigma}$
and $\alpha_{(i_0)}$ is a normalisation constant that is ideally chosen as
\begin{equation}
  \label{eq:7}
  \alpha_{(i_0)} = 1 / \sqrt{\left (\bm{s}^{(i_0)} \right)^{T} \bm{\Sigma}^{-1}
    \bm{s}^{(i_0)}} \; \text{.}
\end{equation}
The matched filter from Eq.~\eqref{eq:6} normalised according to
Eq.~\eqref{eq:7} then defines $\mathcal{T}^{(i_0)}$ in Eq.~\eqref{eq:5} as
multiples of the standard deviation of the filtered noise.  We thus may reject
$\mathcal{H}_0$ and decide on $\mathcal{H}_1$ when $\mathcal{T}^{(i_0)} > \gamma$ and
speak of a line detected at a ``$\gamma \times \sigma$'' significance.
$\mathcal{T}^{(i_0)}$ is therefore also called the \textit{detection significance}.

We note that standard conversion of this ``$\gamma\times\sigma$'' detection
significance into a probability of ruling out $\mathcal{H}_0$
\citep[e.g.,][]{Wall1979}, also known as ``false detection probability'', is
formally only valid if $i_0$ is known a-priori.  While such situations exist in
astronomy, i.e.,  in follow up spectroscopy to detect an emission line of an
already known source \citep[see, e.g.,][]{Loomis2018}, this is not the case in
blind searches for emission lines.  The resulting statistical subtleties for
calculating false detection rates in blind searches on matched filtered data
sets have been extensively studied by \cite{Vio2016}, \cite{Vio2017}, and
\cite{Vio2019} for sub-mm aperture synthesis imaging.  While the details are
involved, generally the real detection significance of an a priori unknown
source is lower than what the standard conversion would suggest.  Perhaps more
intuitively, since $\mathcal{T}^{(i_0)}$ represents the maximised
\textit{signal-to-noise ratio} (SN) of the emission line signal at $i_0$
according to the matched filtering theorem \citep[e.g.,][]{Das1991}, the
hypothesis test can also be regarded as a minimum SN criterion for the detection
of emission lines.  Therefore, $\mathcal{T}^{(i_0)}$ can also be referred to as 
the \textit{SN of an emission line}.

In a blind search for emission lines in $\bm{f}$ we use Eqs.~\eqref{eq:5} --
\eqref{eq:7} to compute the vector
\begin{equation}
  \label{eq:8}
  \bm{\mathcal{T}} = (\mathcal{T}^{(0)}, \dots, \mathcal{T}^{\left(\kappa -1\right)} ,
  \dots, \mathcal{T}^{\left ( N - \kappa + 1 \right)}, \dots,  \mathcal{T}^{(N)})^T 
\end{equation}
and then select the peak (or peaks, if multiple lines are in the spectrum) that
pass the desired threshold.  Only the elements $\mathcal{T}^{(i)}$ with
$i \geq \kappa - 1$ and $i \leq N - \kappa + 1$ can provide a valid test
statistic that is not biased by edge effects.  Explicitly, the computation of
the elements $\mathcal{T}^{(i)}$ of $\bm{\mathcal{T}}$ in Eq.~\eqref{eq:8} is
achieved by replacing $i_0$ with $i$ in Eq.~\eqref{eq:5}:
\begin{equation}
  \label{eq:9}
  \mathcal{T}^{(i)} = \sum_{j = -\kappa}^{\kappa}  T^{(i)}_j f_{i-j} \;\text{.}
\end{equation}
Eq.~\eqref{eq:9} describes the discrete convolution of the of the data vector with the
matched filter.   
The vector $\bm{\mathcal{T}}$ containing the test for each
spectral bin in Eq.~\eqref{eq:8} can thus be computed via
\begin{equation}
  \label{eq:10}
  \bm{\mathcal{T}} = \bm{T} \cdot \bm{f} \; \text{,}
\end{equation}
where $\bm{T}$ is the matrix with elements $T^{(i)}_j$; $j$ indexes the columns
and $i$ indexes the rows of this matrix.  We understand $T^{(i)}_j$ as the
$j$-th element of the matched filter for an emission line source at position $i$
in $\bm{f}$.  Thus, the matrix $\bm{T}$ collects all the filter profiles for
each $i$, and we can view it as a row stack of the vectors $\bm{T}^{(i)}$ from
Eq.~\eqref{eq:6} normalised by Eq.~\eqref{eq:7}.  Importantly, $\bm{T}$ is a
banded sparse matrix, since $T^{(i)}_j = 0$ for $|j| > \kappa$.  Multiplications
with sparse matrices can be computed efficiently, and high-level interfaces to
such efficient implementations exist, e.g., in SciPy's
\texttt{sparse.csr\_matrix} routines \citep{Virtanen2020}.

From now on we assume uncorrelated noise, i.e., that the co-variance matrix is diagonal,
\begin{equation}
  \label{eq:11}
  \Sigma_{ij} = 0 \Leftrightarrow i \neq j \qquad \text{and} \qquad \Sigma_{ii}
  = \sigma_i^2 \;\text{.}
\end{equation}
In this case the elements of $\bm{n}$ (Eq.~\ref{eq:3}) are independent (but not
identically distributed) Gaussian random variables.  Moreover, the inverse of
the co-variance matrix is now
\begin{equation}
  \label{eq:12}
  \bm{\Sigma}^{-1} = \mathbb{1} \cdot (\sigma^{-2}_1, \dots, \sigma^{-2}_N)^T \;\text{,}
\end{equation}
with $\mathbb{1}$ being the identity matrix; $\bm{\Sigma}^{-1}$ is thus a
diagonal matrix with $\Sigma^{-1}_{ij} = 0$ and
$\Sigma_{ii}^{-2} = \sigma_i^{-2}$.  According to Eq.~\eqref{eq:6},
Eq.~\eqref{eq:7}, and Eq.~\eqref{eq:12} then the $T^{(i)}_j$ in
Eq.~\eqref{eq:9}, i.e., the elements of the matrix $\bm{T}$ in
Eq.~\eqref{eq:10}, are
\begin{equation}
  \label{eq:13}
  T^{(i)}_j = \frac{1}{\sqrt{\sum_{j=-\kappa}^{\kappa}
      \frac{s_j^2}{\sigma_{i-j}^2}}} \times \frac{s_j}{\sigma_{i-j}^2} \;\text{.}
\end{equation}

Introducing the voxels $F_{x,y,z}$ of a 3D data cuboid $\bm{F}$ and the
associated voxels $\sigma^2_{x,y,y}$ of the variance data cuboid $\bm{\sigma}$,
both of dimensions $X\times Y\times Z$ ($X$ and $Y$ are the spatial dimensions,
while $Z$ is the spectral dimension), as well as the voxels $s_{x',y',z'}$ of
the normalised 3D line source template $\bm{S}$ (with
$\sum_{i,j,k} S_{i,j,k} \equiv 1$) of dimensions $X'\times Y'\times Z'$, with
$X'\ll X$, $Y'\ll Y$, and $Z'\ll Z$, the expressions in Eq.~\eqref{eq:9} and
Eq.~\eqref{eq:13} are then trivially generalised to 3D:
\begin{equation}
  \label{eq:14}
  \mathcal{T}^{(x,y,z)} =   \sum_{i = -d}^{d} \sum_{j =
    -e}^{e} \sum_{k = -f}^{f}  T^{(x,y,z)}_{i,j,k} F_{x-i,y-j,z-k}   \;\text{,}
\end{equation}
with
\begin{equation}
  \label{eq:15}
  T^{(x,y,z)}_{n,m,l} = \frac{1}{\sqrt{\sum_{i=-d}^{+d} \sum_{j=-e}^{e} \sum_{k=-f}^{f}
      \frac{S_{i,j,k}^2}{\sigma_{x-i,y-j,z-k}^2}}} \times \frac{S_{n,m,l}}{\sigma_{x-n,y-m,z-l}^2}  \;\text{.}
\end{equation}
Without loss of generality here $S_{0,0,0}$ was set as the central pixel of the
line source template, such that the summation indices $i$, $j$, and $k$ cover
the whole extent of $\bm{S}$, i.e.
\begin{equation*}
2(d,e,f) + 1 = (X', Y', Z') \;\text{,}
\end{equation*}
or, equivalently, $S_{i,j,k} \equiv 0$ for $|i| > d$, $|j| > e$, and $|k| > f$.
In the following the limits of the summations will be omitted and $\sum_{i,j,k}$
will be used as a shorthand instead.

Equation~\eqref{eq:14} and Eq.~\eqref{eq:15} define the starting point of the in
\mbox{LSDCat} adopted matched filtering solution for detecting emission lines in
IFS data cuboids; $x$ and $y$ are the spatial coordinates (spaxel coordinates),
and $z$ indexes the cuboid layers in spectral direction (cuboid layer
coordinates).  Equation~\eqref{eq:15} describes the voxels of the 3D matched filter
for the template $\bm{S}$ at position $x,y,z$ in $\bm{F}$.  Moreover,
$S_{0,0,0}$ equates with the peak of the template profile in \mbox{LSDCat}.  In
principle a reformulation of Eq.~\eqref{eq:14} into a matrix-vector dot-product
like Eq.~\eqref{eq:10} is also possible \citep[see Appendix A.3
of][]{Ramos2011}, but for uncorrelated noise this appears not especially
practical and, hence, is not further pursued here.

In analogy to the vector of Eq.~\eqref{eq:8}, we compute with Eq.~\eqref{eq:14}
the voxels of the cuboid $\bm{\mathcal{T}}$, where peaks above desired detection
threshold, $\gamma$, constitute ``$\gamma \times \sigma$'' detections.  In
LSDCat we dub $\bm{\mathcal{T}}$ as ``SN-cuboid'', in reference to the
SN-maximising characteristic of the matched filter, and we baptise $\gamma$ as
SN-threshold $\mathrm{SN}_\mathrm{thresh}$ in the following.

As in the original implementation of LSDCat \citepalias{Herenz2017}, the
template $\bm{S}$ is optimised for emission line sources, whose spatial and
spectral properties are independent:
\begin{equation}
  \label{eq:16}
  \bm{S} = \bm{S}^\text{spat} \otimes\bm{S}^\text{spec} \;\text{,}
\end{equation}
where $\bm{S}^\text{spat}$ denotes the spatial profile (dimensions
$X' \times Y'$), $\bm{S}_\text{spec}$ denotes the spectral profile (dimension
$Z'$), and $\otimes$ denotes the outer product. The voxels of $\bm{S}$ in
Eq.~\eqref{eq:16} are thus given by
\begin{equation}
  \label{eq:17}
  S_{x,y,z} = S^{\text{spat}}_{x,y} \, S^{\text{spec}}_{z}\;\text{.}
\end{equation}
As we will demonstrate below, this separation provides significant benefits for
the computation of $\bm{\mathcal{T}}$.  We provide a brief overview of the
parametric functions adopted for $S^{\text{spat}}_{x,y}$ and
$S^{\text{spec}}_{z}$ in the current version of LSDCat in \ref{sec:temp}.

The use of uncorrelated variances (Eq.~\ref{eq:11}), which allows to write the
matched filter with the simple expressions of Eq.~\eqref{eq:9} or
Eq.~\eqref{eq:15} for the 1D or 3D case, respectively, is born out of practical
necessity.  As explained above, resampling from detector space to the data
cuboid space certainly introduces co-variance terms (see Fig.~5 in
\citealt{Bacon2017} for a visualisation of the spatial correlation in MUSE), but
the inflation in data volume by storing this information is deemed
computationally not tractable in the current wide-field IFS data reduction
pipelines\footnote{Because of the rapid advancements in computer technology,
  this is not seen as a limitation in the future anymore and the pipeline of the
  planned BlueMUSE IFS \citep{Richard2019} will take into account for
  co-variances due to resampling \citep{Weilbacher2022}.}.  This enforces the
design of our current algorithm to work without co-variances.  A side effect of
ignoring the co-variances in the data reduction is, that the absolute values of
$\sigma_{x,y,z}^2$ are underestimated.  Thus, a multiplicative rescaling of the
variance cuboid is required in order to obtain correct variances peak values for
line detections.  Empirical methods for rescaling the variance cuboid exist (see
Sect.~3.1.1 in \citealt{Herenz2017a}, Sect.~3.2.4 in \citealt{Urrutia2018} and
Sect.~3.1.5 in \citealt{Bacon2017}).  Moreover, all formal considerations
regarding the statistical significance of the detections from the SN peaks are
only valid as long as $\bm{S}$ matches the real emission line signal in the
cuboid, but in reality a diversity of emission line shapes will be encountered;
we will deal explicitly with the effects of such mismatches between filter and
emission line signals in the context of MUSE survey data in Sect.~\ref{sec:rs}.

\subsection{The original LSDCat Ansatz and its flaw}
\label{sec:classic}

The evaluation of Eq.~\eqref{eq:14} with $T^{(x,y,z)}_{i,j,k}$ from
Eq.~\eqref{eq:15} is computationally challenging due to the large dimensions of
the MUSE data (typically
$X \times Y \times Z \simeq 300 \times 300 \times 3800 = 3.42 \times 10^8$
voxels, but some studies require mosaiced ``super-cuboids'' with significantly
larger spatial dimensions, see, e.g., \citealt{SanchezAlmeida2022}).  In the
original implementation of LSDCat we thus computed instead
\begin{equation}
  \label{eq:18}
  \mathcal{T}_\text{classic}^{(x,y,z)} = \frac{\sum_{i,j,k}
    S_{i,j,k} F_{x-i,y-j,z-k}}{\sqrt{\sum_{i,j,k} S^2_{i,j,k}
      \sigma^2_{x-i,y-j,z-k}}} \; \text{.}
\end{equation}
With the help of Eq.~\eqref{eq:17} we could separate Eq.~\eqref{eq:18} into two
simple convolution operations, that can be computed quickly via Fast-Fourier
transformation for $\bm{S}^\text{spat}$ and sparse-matrix multiplication for
$\bm{S}^\text{spec}$.

We can justify the use of Eq.~\eqref{eq:18}, by noting that Eq.~\eqref{eq:14}
with the formal matched filter $T^{(x,y,z)}_{i,j,k}$ according to
Eq.~\eqref{eq:15} is well approximated by Eq.~\eqref{eq:18}, provided that
$\sigma^2_{x,y,z}$ does not vary strongly within the limits of summation.  We
demonstrate this for the 1D case, i.e.,  Eq.~\eqref{eq:9} with $T^{(i)}_j$
according to Eq.~\eqref{eq:13}.  Setting $\sigma_i = \sigma_\mathrm{const}$ in
Eq.~\eqref{eq:9} and Eq.~\eqref{eq:13} we find
\begin{equation}
  \label{eq:19}
  \mathcal{T}^{(i)} \approx \frac{\frac{1}{\sigma_\text{const.}^2} \sum_j s_j
    f_{i-j}}{\frac{1}{\sigma_\text{const.}} \sqrt{\sum_j s_j^2}} = \frac{\sum_j s_j
    f_{i-j}}{\sigma_\text{const} \sqrt{\sum_j s_j^2}} \;\text{.}
\end{equation}
And for the 1D-case of Eq.~\eqref{eq:18},
\begin{equation}
  \label{eq:20}
  \mathcal{T}^{(i)}_\text{classic} = \frac{\sum_j s_j f_{i-j}}{\sqrt{\sum_j s_j^2
    \sigma_{i-j}^2 }} \; \text{,}
\end{equation}
we have with $\sigma_i := \sigma_\mathrm{const}$:
\begin{equation}
  \label{eq:21}
  \mathcal{T}^{(i)}_\text{classic} \approx \frac{\sum_j s_j
    f_{i-j}}{\sigma_\text{const} \sqrt{\sum_j s_j^2}} \;\text{.}
\end{equation}
This is identical to Eq.~\eqref{eq:19}.  However, when there are strong
variations of the variance values within the summation limits, then
Eq.~\eqref{eq:20}, or Eq.~\eqref{eq:18} for the 3D case, does not produce
correct results.   In this case we expect the detection significances
to be biased towards lower values.

\begin{figure}
  \centering
  \includegraphics[width=0.5\textwidth]{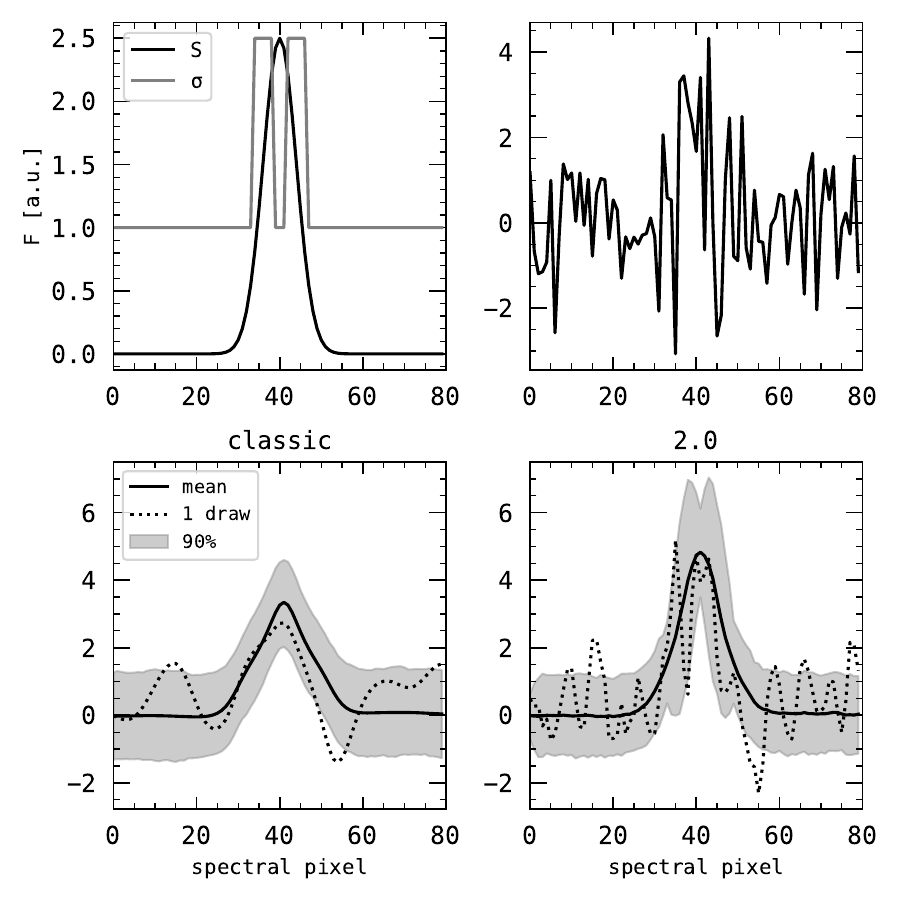}
  \caption{
    Illustration of the difference in SN between a simplified matched filter
    that assumes variances to be constant over the filter profile and the
    correct matched filter that correctly accounts for uncorrelated variances.
    the The \textit{top left} panel displays the ground truth: an emission line
    signal $\bm{s}$ (black line) and a corresponding variance vector (we show
    the square-root $\sigma$ as grey line).  The \textit{top right} panel shows
    a Monte-Carlo realisation of a vector $\bm{f}$ resulting from this set-up.
    The \textit{bottom left} panel shows the SN spectrum from the approximate
    matched filter ($\mathcal{T}^{(i)}_\mathrm{classic}$; Eq.~\ref{eq:20}),
    whereas the \textit{bottom right} panel shows the SN spectrum $\bm{T}$ from
    the correct matched ($\mathcal{T}^{(i)}$; Eq.~\ref{eq:9} with
    Eq.~\ref{eq:13}).  In both bottom panels we show the average from 1000
    realisations (\textit{black line}), the 90 percentile (\textit{grey shaded
      region}) and the SN spectrum of the example realisation from the top right
    panel (dotted line).  }
  \label{fig:mfdemo}
\end{figure}

We illustrate the difference between $\mathcal{T}^{(i)}_\mathrm{classic}$
(Eq.~\ref{eq:20}) and $\mathcal{T}^{(i)}$ (Eq.~\ref{eq:9} with Eq.~\ref{eq:13})
in Figure~\ref{fig:mfdemo} for an example of increased variances in the wings of
an idealised 1D emission line signal that is recovered with its matched filter.
Here the signal is assumed to be a perfect 1D Gaussian (dispersion = 4 bins;
peak amplitude = 2.5). For each bin a unit variance is assumed, except for 4
bins leftwards and rightwards of the peak position, where a variance of 6.25 has
been assumed; the total length of $\bm{f}$ is 81 bins and the peak of the line
is placed at $i_0 = 40$.  The ground truth according to this set-up is shown in
the upper left panel of Figure~\ref{fig:mfdemo}.  A particular realisation of
this set-up is shown in the upper right panel of this figure, where the line
signal clearly got buried in noise.  In the bottom two panels of
Figure~\ref{fig:mfdemo} we contrast the resulting vectors $\bm{\mathcal{T}}$
(Eq.~\ref{eq:8}) where the $\mathcal{T}^{(i)}$ have been calculated with the
classic formalism (Eq.~\ref{eq:20}; bottom left panel) and the correct matched
filter formalism (Eq.~\ref{eq:9} with Eq.~\ref{eq:13}; bottom right panel).
These results are from $10^3$ Monte-Carlo realisations of the set-up.  It can be
seen that the peak values $\mathcal{T}^{(i_0)}$, which is the relevant decisions
statistic for the presence or absence of a source, are clearly lower for the
simplified formulae ($\mathcal{T}^{(i_0)}_\mathrm{classic}$ -- mean: 3.3;
10-/90-percentile: 2.0/4.5) then for the proper matched-filter (mean: 4.8;
10-/90-percentile: 3.3/6.4).  Or, putting it differently, in 72\% of all
detection attempts with $SN_\mathrm{thresh} = 4 $ the line would have been
rejected with the classic LSDCat Ansatz, whereas this happens only for 12\% of
all attempts with the same threshold in the correct formalism.

While the idealised set-up in Figure~\ref{fig:mfdemo} is contrived for
illustrative purposes, spectrally rapidly varying variances of considerable
amplitude appear in the red part of the optical spectral range (R and I-band)
due to the air glow background \citep[see Sect.~4 of][]{Noll2012}.  Here
predominantly the molecular Meinel OH-bands are contributing, which appear like
combs in 1D spectral representations \citep[see also Figures
in][]{Hanuschik2003} and the spectral distance between individual air glow lines
is only a few \AA{}.  The result is that emission line signals which are
overlapping with the OH-lines will not be optimally recovered by the classic
algorithm.

\subsection{The improved Ansatz (LSDCat\,2.0)}
\label{sec:modified}

Our goal for the improved algorithm is now to account for spectrally varying
background in IFS data cuboids while maintaining the computational simplicity of
the original algorithm.  This can be achieved, when we assume that the variances
are not varying as a function of position, i.e.,  when we set
\begin{equation}
  \label{eq:22}
  \sigma_{x,y,z}^2 := \sigma_{z}^2 \;\text{.}
\end{equation}
Under this condition we can separate the template in Eq.~\eqref{eq:15} using
Eq.~\eqref{eq:17}, i.e., 
\begin{equation}
  \label{eq:23}
  T^{(x,y,z)}_{i,j,k} = T_k^{\{z\}} T_{ij}^{\{z\}} \;\text{,}
\end{equation}
with
\begin{equation}
  \label{eq:24}
  T_k^{\{z\}} = \frac{1}{\sqrt{\sum_n
      \frac{\left (S_n^{\{z\}} \right)^2}{\sigma^2_{z-n}}}} \times \frac{S^{\{z\}}_k}{\sigma_{z-k}^2}
\end{equation}
and
\begin{equation}
  \label{eq:25}
  T_{ij}^{\{z\}} = \frac{S_{ij}^{\{z\}}}{\sqrt{\sum_{lm}
      \left (S_{lm}^{\{z\}} \right )^2}}  \;\text{.}
\end{equation}
The superscript ${\{z\}}$ in Eqs.~\eqref{eq:23} -- \eqref{eq:25} is meant to
indicate the implicit dependence of our templates $\bm{S}^\text{spat}$ and
$\bm{S}^\text{spec}$ on wavelength and thus on the index $z$ in spectral
direction (see \ref{sec:temp} for the rationale).  We note that the spatial
filter in Eq.~\eqref{eq:25} is just the re-normalised spatial template and the
spectral filter in Eq.~\eqref{eq:24} is identical to the spectral filter in the
1D case introduced in Sect.~\ref{sec:mf} (Eq.~\ref{eq:13}).  Plugging
Eq.~\eqref{eq:23} into Eq.~\eqref{eq:14} we find that the voxels of the
SN-cuboid can be computed via
\begin{equation}
  \label{eq:26}
  \mathcal{T}^{(x,y,z)} = \sum_k T_k^{\{z\}} \left ( \sum_{ij} T_{ij}^{\{z\}}
    F_{x-i,y-j,z-k}  \right) \;\text{.}
\end{equation}

The numerical computation of Eq.~\eqref{eq:26} can be carried out efficiently.
First, we evaluate Eq.~\eqref{eq:25} for all $z$ for the parametric models for
$S_{ij}^{\{z\}}$ (Eq.~\ref{eq:66} or Eq.~\ref{eq:67} in \ref{sec:temp}) and we
then use the discrete fast Fourier transform to calculate the convolution sum in
brackets of Eq.~\eqref{eq:26} for each $z$.  This operation can be trivially
parallelised by partitioning $\bm{F}$ along the spectral axis into 
sub-cuboids for each CPU core.  For the outer sum we first pre-compute all
$T_k^{\{z\}}$ according to Eq.~\eqref{eq:24} for the given spectral template
(Eq.~\ref{eq:69} in \ref{sec:temp}).  Then we see the similarity between the
outer sum and Eq.~\eqref{eq:9}, i.e., we can compute for each spaxel a (sparse)
matrix - vector product according to Eq.~\eqref{eq:10}.  Again, this step is
trivially parallelised, now by partitioning the data cuboid over the spatial
axes.

Above simplification only works under the provision that Eq.~\eqref{eq:22}
holds, i.e.,  that the variances do not vary as a function of position.  We found
that this is indeed the case for MUSE data cuboids obtained from observations
that follow the recommended dithering and rotation strategies.  In fact,
\cite{Herenz2017a}, when compiling the first catalogue of emission line galaxies
in the MUSE-Wide survey, already calculated an 1D empirical variance spectrum
$\sigma^2_z$, that was then blown up to a variance cuboid via
$\sigma^2_{x,y,z} = N^\mathrm{exp}_{x,y}/N_\mathrm{max} \cdot \sigma^2_z$, where
$N^\mathrm{exp}_{x,y}$ denotes the exposure count map that counts the average
number of exposures that contribute to each spaxel and $N_\mathrm{max}$ denotes
the maximum number of exposures, i.e., 
$N_\mathrm{max} = \max_{x,y} N^\mathrm{exp}_{x,y}$.  The procedure to calculate
such an empirical variance spectrum was then refined by \cite{Urrutia2018}, but
again $N^\mathrm{exp}_{x,y}$ was used for rescaling.  Significant differences of
$N^\mathrm{exp}_{x,y}$ with respect to $N_\mathrm{max}$ occur near the edges of
the field of view, but here edge effects render the decision statistic from the
matched filter unreliable anyway.

\subsection{Verification of the improved method}
\label{sec:verify}

\begin{figure*}[t!]
  \centering
  \includegraphics[width=0.85\textwidth, trim=10 10 20
  10,clip=True]{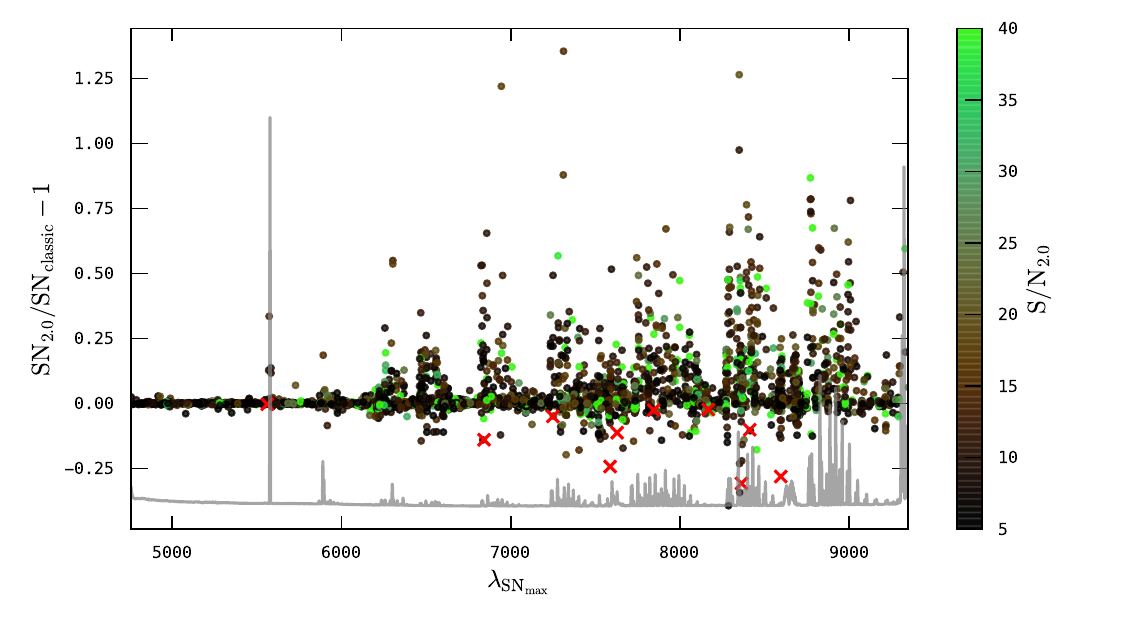}
  \caption{ Comparison between of the SNs obtained with the improved filter,
    $\mathrm{SN}_{2.0}$ (Eq.~\ref{eq:26}), to those obtained with the classic
    filter, $\mathrm{SN}_\mathrm{classic}$ (Eq.~\ref{eq:18}), for all
    catalogued emission lines of the MUSE-Wide data release by
    \cite{Urrutia2018}.  The relative difference of $\mathrm{SN}_{2.0}$ with
    respect to $\mathrm{SN}_\mathrm{classic}$ is shown as a function of peak
    detection wavelength $\lambda_\mathrm{SN_\mathrm{peak}}$, and the absolute
    $\mathrm{SN}_{2.0}$ is colour coded as indicated by the colourbar
    (detections with $\mathrm{SN}_{2.0} > 40$ are not assigned to a different
    colour than $\mathrm{SN}_{2.0} = 40$ detections).  The 10 emission lines
    for which $\mathrm{SN}_{2.0}$ is below the detection threshold of the
    original catalogue are marked by red crosses.  An arbitrarily scaled
    variance spectrum is plotted as a grey line to indicate the spectral
    variation of the telluric background. }
  \label{fig:comp}
\end{figure*}

\begin{figure}[t]
  \centering
  \includegraphics[width=0.45\textwidth,trim=20 10 20
  0,clip=True]{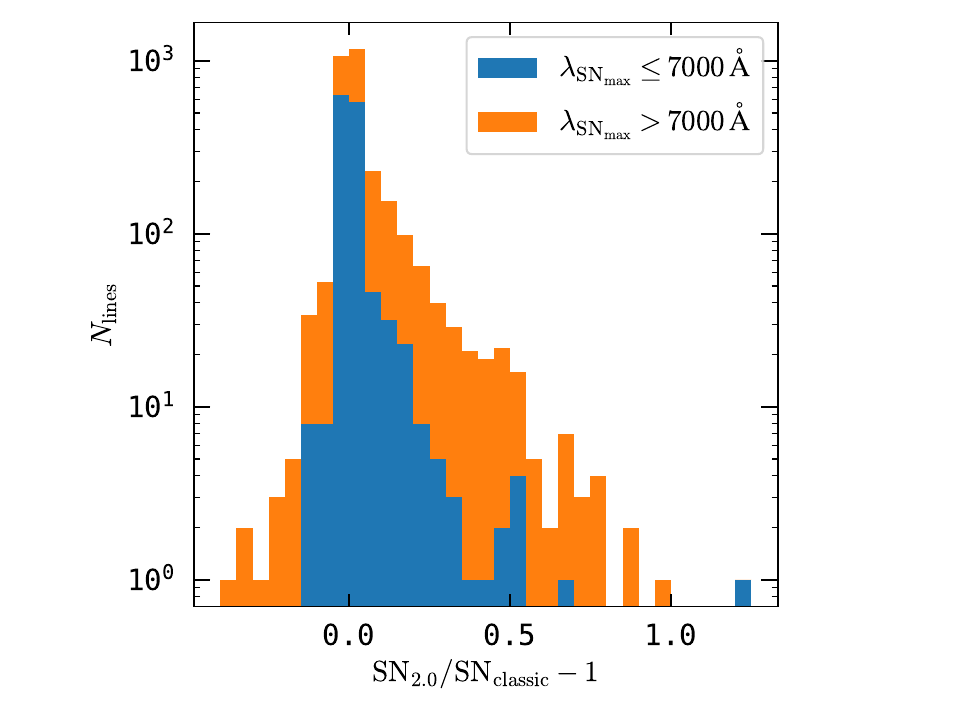}
  \caption{Stacked histogram (bin width = 0.05) of the relative differences of
    $\mathrm{SN}_{2.0}$ with respect to $\mathrm{SN}_\mathrm{classic}$
    (cf. Fig.~\ref{fig:comp}); emission lines at
    $\lambda_\mathrm{SN_\mathrm{peak}} \leq 7000$\,\AA{} are counted in blue,
    those at $\lambda_\mathrm{SN_\mathrm{peak}} 7000$\,\AA{} are counted in
    orange.  }
  \label{fig:hist}
\end{figure}

We verified the expected improvement of detection significances with the method
presented in Sect.~\ref{sec:modified} by a reanalysis of the publicly available
data of MUSE-Wide survey \citep{Urrutia2018}.  To this aim we reprocessed all
released data cuboids in the same way as in the \citeauthor{Urrutia2018} study,
i.e.,  we used the template parameters provided in Table~2 of \cite{Urrutia2018},
but now with the \mbox{LSDCat2.0} routines.  For each detected emission line we
then compared the peak SN values between the original Ansatz
(Sect.~\ref{sec:classic}) and the improved Ansatz (Sect.~\ref{sec:modified}) for
all 3057 emission lines in the \citeauthor{Urrutia2018} catalogue.  The result
of this exercise is presented in Figure~\ref{fig:comp}, where we show for each
catalogued detection the relative difference between the new and the old method,
$\mathrm{SN}_{2.0}/\mathrm{SN}_\mathrm{classic} - 1$, as a function of detection
wavelength.  Here $\mathrm{SN}_{2.0}$ refers to the peak values from
Eq.~\eqref{eq:26}, while $\mathrm{SN}_\mathrm{classic}$ refers to the peak
values according to Eq.~\eqref{eq:18}.  We also inset an arbitrarily scaled
effective variance spectrum into Figure~\ref{fig:comp} to illustrate the
position and amplitude of the air-glow lines in the MUSE spectral range.

As can be seen from Figure~\ref{fig:comp}, the improved Ansatz indeed boosts the
SN of many detected emission lines especially in the forest of air glow lines.
The overall improvement also becomes apparent in the histogram of relative SN
differences shown in Figure~\ref{fig:hist}.  There we also separated the counts
in each bin by their spectral position in the data cuboid, and again it can be
seen how, especially in the red part of the spectrum, a significant number of
sources experiences a boost in their detection significances.  In numbers: 1892
detections have $\mathrm{SN}_{2.0} > \mathrm{SN}_\mathrm{classic}$, whereas for
1165 lines $\mathrm{SN}_{2.0} < \mathrm{SN}_\mathrm{classic}$.  However, for the
vast majority of detections where the improved algorithm provides a lower SN,
the difference is marginal, whereas the boost for the former can be substantial.
For example, while only 99 detections have
$\mathrm{SN}_{2.0} < 0.95 \times \mathrm{SN}_\mathrm{classic}$, 722 show
$\mathrm{SN}_{2.0} > 1.05 \times \mathrm{SN}_\mathrm{classic}$.

Nevertheless, we also find in this comparison that ten emission lines show a
peak value that is marginally lower than the original detection threshold.  All
these catalogued lines are overlapping with air-glow lines, one even with the
extremely strong [\ion{O}{I}] $\lambda5577$ line.  For those lines the
$\mathrm{SN}_\mathrm{classic}$ value got artificially boosted due to a strong
positive noise contribution that gets adequately suppressed in the new
formalism.  Eight of those ten lines are secondary lines of emission line
galaxies that have been detected in other, stronger, lines, i.e., these galaxies
are not lost from the catalogue.  The remaining two detections\footnote{ID
  129026144 and 146018244.} were supposed Ly$\alpha$ emitters, but these low SN
detections were assigned also with the lowest confidence (i.e., $\approx$50\%
probability of being false detections according to \citealt{Urrutia2018}).
Thus, our new algorithm does not remove significant detections from the previous
catalogue, but significantly improves the detection significances of emission
lines that are overlapping with air-glow lines.  Hence, this little experiment
verifies indeed that the new algorithm performs as expected.

We limited ourselves here to only check emission lines that were already
catalogued.  Nevertheless, because of the overall improvement of low-SN
detections within the forest of the OH-bands we foresee that the
\mbox{LSDCat2.0} algorithm also uncovers new lines in those spectral regions.
However, identifying and classifying new detections still requires manual
inspection and verification by a team of experts (see \citealt{Inami2017} and
\citealt{Urrutia2018} for descriptions of this process).  Such a complete
reanalysis of the MUSE-Wide dataset, clearly beyond the scope of the present
work, may be desired for subsequent data MUSE-Wide releases.

\section{The selection function of the 3D matched filter}
\label{sec:selfun}

A selection function, $f_C$ (sometimes also referred to as completeness
function, e.g., in \citealt{Caditz2016}), encodes the probability that a source
with given attributes will enter our catalogue.  Selection functions are needed
whenever we want to construct a model of reality given the catalogue data
\citep[see][]{Rix2021}.  Relevant applications in the context of surveys for
emission line objects with IFS are, e.g., galaxy luminosity functions or
modelling the galaxy power spectrum via correlation analysis
\citep[e.g.,][]{Blanc2011,Herenz2019,HerreroAlonso2021,Zhang2021}.  In terms of emission
line source detection in IFS data cuboids the most important parameters are the
wavelength, $\lambda$, of the detection and the flux of the emission line,
$F_\mathrm{line}$.

Processing the flux data cuboids with a 3D matched filter (Eq.~\ref{eq:14} or,
more specifically, Eq.~\ref{eq:26}), can be understood as a linear mapping of
$F_\mathrm{line}$ to SN space: $SN \propto F_\mathrm{line}$ (we recall from
Sect.~\ref{sec:mf} that $\mathcal{T}^{x,y,z}$ can be understood as the maximised
SN for a emission line at position $x,y,z$ in the data cuboid).  The
proportionality factor depends on the noise properties encountered at the
spatial ($x,y$) and spectral ($\lambda$ or $z$) position of the emission line
source in the data cuboid and on the level of congruence between template
$\bm{S}$ and actual emission line source.  For the construction of the filter we
explicitly assumed in Sect.~\ref{sec:modified} that the noise properties are not
an implicit function of position ($x,y$) and that the effect of varying depth
over the field of view can be accounted for by rescaling.  Moreover, because of
the randomness of the noise, the value of SN is stochastic and thus we can
abbreviate the mapping as
\begin{equation}
  \label{eq:27}
  E[SN_\mathrm{\lambda}] = C(\lambda) \times F_\mathrm{line}\;\text{,}
\end{equation}
where $E[SN_\mathrm{\lambda}]$ denotes the expectation value of
$SN_\mathrm{\lambda}$ and $C(\lambda)$ is in units of inverse flux.  We also
assumed that the noise is normally distributed and uncorrelated, thus the
distribution of $SN_\mathrm{\lambda}$ for a given $F_\mathrm{line}$,
$g(SN_\lambda | F_\mathrm{line})$, is also normal and with variance
$V[SN_\mathrm{\lambda}] = 1$, i.e., 
\begin{equation}
  \label{eq:28}
  g(SN_\lambda | F_\mathrm{line}) = \frac{1}{\sqrt{2}} \exp \left ( - \frac{(SN_\lambda -
      C(\lambda) F_\mathrm{line})^2}{2} \right ) \;\mathrm{.}
\end{equation}
Now the selection function,
$f_C(F_\mathrm{line}, \lambda | \mathrm{SN}_\mathrm{thresh}) \in [0,1]$,
encodes the cumulative probability that $F_\mathrm{line}$ will result in
$SN_\mathrm{\lambda}(F_\mathrm{line}) > \mathrm{SN}_\mathrm{thresh}$, i.e., 
\begin{equation}
  \label{eq:29}
  f_C(F_\mathrm{line}, \lambda | \mathrm{SN}_\mathrm{thresh}) =
  \int_{\mathrm{SN}_\mathrm{thresh}}^\infty g(SN_\lambda | F_\mathrm{line})
  \,\mathrm{d}(SN_\mathrm{\lambda}) 
\end{equation}
for a given a detection threshold $\mathrm{SN}_\mathrm{thresh}$.  Using the
error-function,
\begin{equation}
  \label{eq:30}
  \erf (x) = \frac{2}{\sqrt{\pi}}\int_{0}^{x}e^{-t^{2}}\,\mathrm{d}t = 1 -
\frac{2}{\sqrt{\pi}} \int_x^\infty e^{-t^2} \, \mathrm{d} t \;\text{,}
\end{equation}
we can express the selection function in Eq.~\eqref{eq:29} as
\begin{multline}
  \label{eq:31}
  f_C(F_\mathrm{line}, \lambda | \mathrm{SN}_\mathrm{thresh}) 
  = \\ = \frac{1}{2}
  \left [  1 + \erf \left ( \frac{C(\lambda) \times F_\mathrm{line} -
        \mathrm{SN}_\mathrm{thresh}}{\sqrt{2}}  \right )  \right ] \;\text{.}
\end{multline}
Due to the symmetry of the normal distribution, Eq.~\eqref{eq:31} is also
symmetric around $F_\mathrm{line}$, and this flux thus equates with the 50\,\%
completeness limit which we now denote $F_{50}$, i.e., 
$f_C(F_{50}, \lambda | \mathrm{SN}_\mathrm{thresh}) \equiv 0.5$ and
\begin{equation}
  \label{eq:32}
  F_{50} = \frac{\mathrm{SN}_\mathrm{thresh}}{C(\lambda)}
  \;\text{.}
\end{equation}
As can be seen, the selection function is completely determined if the
proportionality factor $C(\lambda)$ is known and in the remainder of this
section we have to deal with calculations of $C(\lambda)$.

\subsection{The idealised selection function}
\label{sec:ideal-select-funct}

We define the idealised selection function as the selection function for an
emission line source in the data cuboid that is perfectly matched by the source
template $\bm{S}$ (Eq.~\ref{eq:16} and Eq.~\ref{eq:17}).  For the calculation of
$C(\lambda)$ we calculate the response of the filter for such a source at some
position $x',y',z'$ in the data cuboid $\bm{F}$:
\begin{equation}
  \label{eq:33}
  F_{x,y,z} = A \cdot S_{x-x',y-y'}^\mathrm{spat} S^\mathrm{spec}_{z-z'} + n_{x,y,z}\;\text{,}
\end{equation}
where $A$ is the amplitude at which the normalised source profile and
$n_{x,y,z}$ describes a source-free noise cuboid.  Then the matched filter of
Eq.~\eqref{eq:26} with the template according to Eq.~\eqref{eq:24} and
Eq.~\eqref{eq:25} has the expectation value
\begin{equation}
  \label{eq:34}
  E_\mathcal{I} \left [ \mathcal{T}^{(x',y',z')} \right ] = A \cdot \sqrt{ \sum_{ij} \left ( S_{ij}^{\{z' \}}
    \right )^2 } \cdot \sqrt{\sum_{k}  \frac{\left ( S_k^{\{z' \}}  \right
      )^2}{\sigma_{z'-k}^2} } \; \text{,}
\end{equation}
where we used $E[n_{x,y,z}] = 0$.  In anticipation of the discussion in
Sect.~\ref{sec:rs} below we wrote $E_\mathcal{I}$ in Eq.~\eqref{eq:34} to
signify that this is the expectation value for the idealised scenario.

A voxel in the flux data cuboid $F_{x,y,z}$ is assumed to encode flux density.
For example, MUSE data reduction pipeline provides fluxes in
erg\,s$^{-1}$cm$^{-2}$\AA{}.  The amplitude A in Eq.~\eqref{eq:33} then relates
to the emission line flux $F_\mathrm{line}$ in erg\,s$^{-1}$cm$^{-2}$ via
\begin{equation}
  \label{eq:35}
A = F_\mathrm{line}/\Delta \lambda \;\text{,}  
\end{equation}
where $\Delta \lambda$ is the native wavelength sampling of the data cuboid
(default $\Delta \lambda = 1.25$\,\AA{} for MUSE).  The wavelength $\lambda$
relates linearly to the spectral coordinate $z$ (cf. Eq.~\ref{eq:70} in
\ref{sec:temp}) and we abbreviate it by $\lambda[z]$.  We now equate
Eq.~\eqref{eq:27} with Eq.~\eqref{eq:34} to find
\begin{equation}
  \label{eq:36}
  C_\mathcal{I}(\lambda [z]) = \frac{1}{\Delta\lambda}  \cdot 
  \sqrt{ \sum_{ij} \left ( S_{ij}^{\{z \}}
    \right )^2 } \cdot \sqrt{\sum_{k}  \frac{\left ( S_k^{\{z \}}  \right
      )^2}{\sigma_{z-k}^2} } \;\text{,}
\end{equation}
where again the subscript $\mathcal{I}$ signifies that this is the
proportionality constant for the idealised selection function.

Equation~\eqref{eq:31} with $C(\lambda[z])$ from Eq.~\eqref{eq:36} provides an
expression for the idealised selection function for the line search with some
template $\bm{S}$ (Eq.~\ref{eq:16}) given an effective variance spectrum
$\sigma^2_z$.

\subsection{Verification of the analytical expression for the idealised
  selection function}
\label{sec:verif-analyt-expr}

\begin{figure}
  \centering
  \includegraphics[width=0.45\textwidth,trim=10 0 3 3
  ,clip=True]{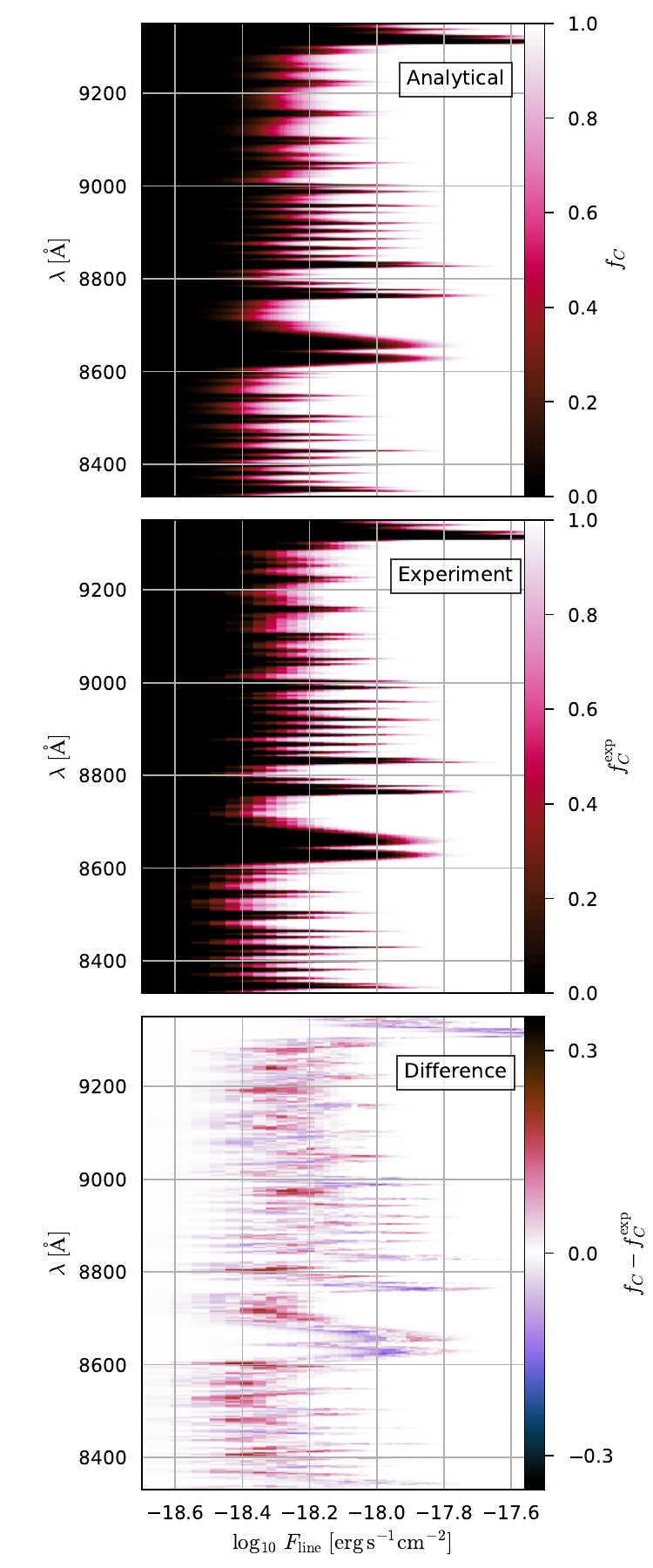} \vspace{-0.5em}
  \caption{Comparison between the idealised selection function according to
    Eq.~\eqref{eq:31} with $C(\lambda)$ from Eq.~\eqref{eq:36} ($f_C$,
    \textit{top panel}) to the idealised selection function from a source
    insertion and recovery experiment ($f_C^\mathrm{exp}$, \textit{centre
      panel}) for a $z > 6$ LAE search with LSDCat2.0 in the MXDF \citep{Bacon2022}
    using the template given in
    Table~\ref{tab:sn2} and a detection threshold of
    $SN_\mathrm{thresh} = 6.41$. The \textit{bottom panel} shows the difference
    $f_C - f_C^\mathrm{exp}$. }
  \label{fig:2dsel}
\end{figure}

\begin{figure}
  \centering
  \includegraphics[width=0.5\textwidth]{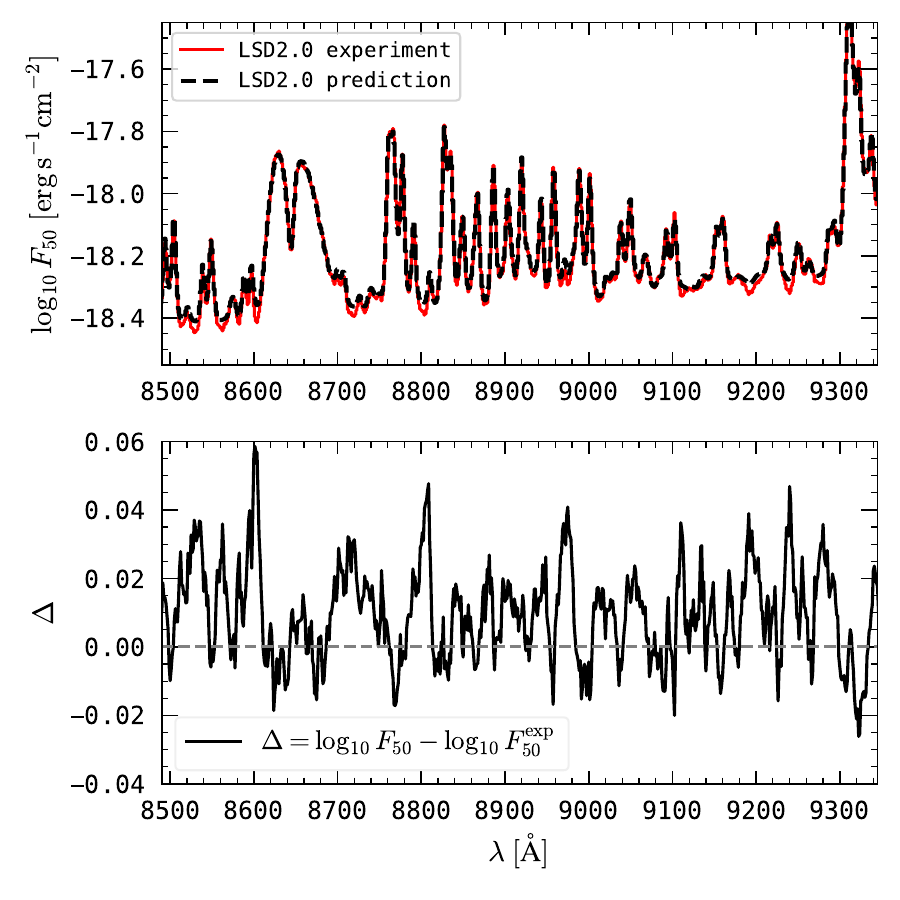}
  \caption{Analytical 50\% completeness estimate in comparison to the 50\%
    completeness achieved in a source insertion and recovery experiment,
    $F_{50}^\mathrm{exp}$, for the LSDCat2.0 line search at
    $\lambda > 8500$\,\AA{} in the MXDF with $SN_\mathrm{thresh} = 6.41$ under
    the assumption of idealised sources described by the 3D template in
    Table~\ref{tab:sn2}.  The \textit{top panel} displays both the analytical
    $\log_{10} F_{50}$ (Eq.~\ref{eq:32} with $C(\lambda)$ from Eq.~\ref{eq:36};
    \textit{dashed black line}) and the $\log_{10} F_{50}^\mathrm{exp}$ from the
    numerical experiment (\textit{red line}), whereas the \textit{bottom panel}
    displays the difference $\Delta = \log_{10} F_{50} - \log_{10} F_{50}^\mathrm{exp}$. }
  \label{fig:f50test}
\end{figure}

\begin{figure*}
  \centering
  \includegraphics[width=\textwidth,trim=30 11 45 0,clip=True]{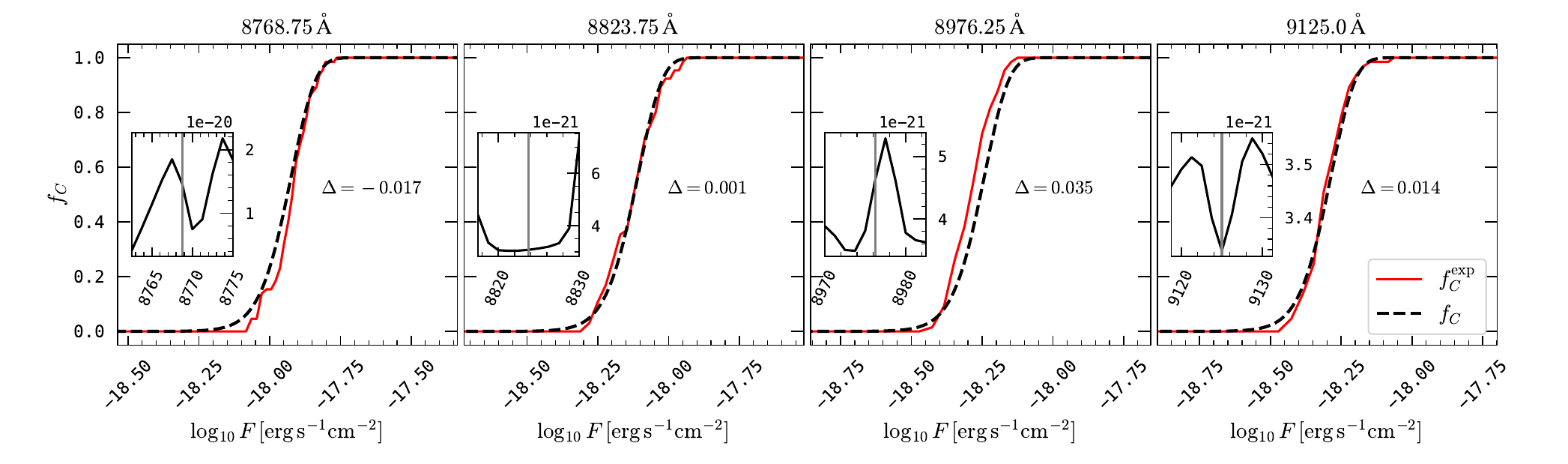}
  \caption{Analytical completeness curves (\textit{dashed black lines};
    Eq.~\ref{eq:31} with $C(\lambda)$ from Eq.~\ref{eq:36}) in comparison to
    the completeness curves from a source insertion and recovery experiment (red
    lines) for the \mbox{LSDCat2.0} line search in the MXDF (see text and
    captions to Figures~\ref{fig:2dsel} and \ref{fig:f50test} for details) at
    four representative wavelength layers.  In each panel we print the
    difference $ \Delta = \log_{10} F_{50}^\mathrm{ana} - \log_{10} F_{50}^\mathrm{exp}$
    (cf. Fig.~\ref{fig:f50test}) and we inset a zoom-in of the extracted average
    noise spectrum ($\sqrt{\sigma^2_{z}}$ in erg\,s$^{-1}$cm$^{-2}$\AA{}$^{-1}$) around the layers of interest.}
  \label{fig:f50crvs}
\end{figure*}

We performed a source insertion and recovery experiment to check whether
Eq.~\eqref{eq:31} with $C(\lambda)$ from Eq.~\eqref{eq:36} can indeed predict
the idealised selection function of the algorithm.  For this experiment we used
the recently released data cuboid of the MUSE eXtreme Deep Field (MXDF) by
\cite{Bacon2022}.  This dataset is the deepest blank field ever obtained with
MUSE ($t_\mathrm{exp} = 141$\,h).  In order to reduce potential systematics,
this programme used a non-standard observing strategy where each subsequent
observing block was rotated by a few degrees.  The result is that the final
combined field of view within the data cuboid is circular.  As groundwork for
future population studies of unprecedented faint Ly$\alpha$ emitters in this
dataset we constructed an emission line emitter catalogue with LSDCat.  For
reference we list the template parameters that were used in this search in
\ref{sec:templ-param-used} (Table~\ref{tab:sn2}).  The detection threshold for
this search is $SN_\mathrm{thresh} = 6.41$.  This value\footnote{The value
  presented here was initially calculated on an internal MUSE consortium data
  release in SN steps of 0.1 (L. Wisotzki, priv. comm.).  The public data
  release then fixed a bug with the variance datacube that required a rescaling
  of the initial detection threshold to a value with two decimal places.} was
found to be the ``point of diminishing returns'' from the ratio of the number of
detections in the normal datacube to the number of detections in the negated
datacube (see Sect.~4.5 in \citetalias{Herenz2017}).  For the demonstration here
we only consider the extreme red of the MUSE spectral range, where the density
and amplitude of the telluric background emission lines is highest.

For the recovery experiment we implanted emission line sources into the MXDF
that are exactly described by this template.  We used a regular grid of 70
insertion positions and we considered only the central region of $> 140$\,h
depth \citep[see Fig.~3 in][]{Bacon2022}.  The distance between neighbouring sources
was 6\arcsec{}.  At each wavelength layer $\lambda[z]$ we then inserted the
artificial line sources with fluxes from
$\log_{10} F_\mathrm{line}^\mathrm{ins}$[erg\,s$^{-1}$cm$^{-2}$]$ = - 18.6$ to
$\log_{10} F_\mathrm{line}^\mathrm{ins}$[erg\,s$^{-1}$cm$^{-2}$]$= -17.1$ at an
increment of $3 \times 10^{-20}$erg\,s$^{-1}$cm$^{-2}$; in each insertion
experiment all sources have the same flux.  After each insertion we ran the full
source detection chain that was used for the construction of the catalogue,
i.e.,  we subtracted a running median to remove continuum emission and we ran the
3D matched-filter with the template given in Table~\ref{tab:sn2}.  We then
counted the number of recovered lines above $SN_\mathrm{thresh}$ which, divided
by the number of inserted sources, provides us with the experimentally
determined completeness
$f_C^\mathrm{exp.}(F_\mathrm{line}^\mathrm{ins}, \lambda[z])$ for each
($F_\mathrm{line}^\mathrm{ins}, \lambda[z]$)-pair.  We excluded detections of
real sources in our tally by excluding insertion positions that overlapped with
detections of real emission line sources in the original search, i.e.,  in some
layers the denominator of inserted sources was less than 70 for the calculation
of $f_C^\mathrm{exp.}(F_\mathrm{line}^\mathrm{ins}, \lambda[z])$.

In Figure~\ref{fig:2dsel} we show a comparison between the analytic prediction
for the idealised case, i.e.,
$f_C(\lambda,F_\mathrm{line} | SN_\mathrm{thresh} = 6.41)$ according to
Eq.~\eqref{eq:31} with $C(\lambda)$ from Eq.~\eqref{eq:36} for the template
parameters given in Table~\ref{tab:sn2}, and the experimentally determined
selection function, $f_C^\mathrm{exp}$, from the above described insertion and
recovery procedure. To compute the analytic prediction we evaluated
Eq.~\eqref{eq:36} we calculated the effective variance spectrum
(Eq.~\ref{eq:22}) as the median in each layer of the variance cube in the
central $141$\,h deep region.  The visual comparison between the analytical
prediction (top panel of Figure~\ref{fig:2dsel}) with the experimental result
(middle panel of Figure~\ref{fig:2dsel}) does not reveal large discrepancies.
Only the sampling in $\log_{10} F$ of the numerical experiment becomes apparent
as pixelated structure in the middle panel.  Obviously, the analytic prediction
can be evaluated on a much finer flux grid than what would be feasible for an
insertion and recovery experiment.  The impression of congruity between
prediction and experiment is verified in the bottom panel of
Fig.~\ref{fig:2dsel}, where we show the difference between the $f_C$ and
$f_C^\mathrm{exp}$.  Only very few $(F_\mathrm{line},\lambda)$-bins show a
difference greater than 20\%, and generally the experimental and analytical
curves are close to each other.

The congruence between prediction and experiment can also be appreciated in
Figure~\ref{fig:f50test}, where we compare the analytic 50\% completeness
(Eq.~\ref{eq:32}) to the 50\% completeness from the experiment.  The latter was
calculated via linear interpolation using the two flux insertion values for
which $f_C^\mathrm{exp}$ had the smallest absolute difference to $0.5$.  Again,
both curves are nearly identical, with the average
$\Delta = \log_{10} F_{50} - \log_{10} F_{50}^\mathrm{exp}$ being $0.01$.
Inspecting Figures~\ref{fig:2dsel} and \ref{fig:f50test} in detail reveals that
the experiment results in slightly deeper flux limits, but the discrepancies are
marginal and dwarfed in comparison to the uncertainties on the selection
function due to mismatches between filter template and real sources (see
Sect.~\ref{sec:rs} below).  Importantly, the fluctuations on $\Delta$ appear
random and do not correlate with features in the sky background.

Lastly, we compare in Figure~\ref{fig:f50crvs} the analytic
$f_C(F_\mathrm{line}, \lambda)$ curves with their experimental counterparts for
four different wavelength layers.  These layers are chosen to be representative
for the typical situations in the background noise spectrum.  The first and the
third panel (from the left) demonstrate the selection function for wavelength
layers that are in the wings of a [OH]-line, whereas the second panel shows a
layer were the surrounding noise spectrum is not varying strongly.  The fourth
panel shows the $f_C$ curve for a layer where the background noise is lower, but
with sharp spikes rising in the redward and blueward layers; this situation is
akin to the example discussed in Sect.~\ref{sec:classic}
(Figure~\ref{fig:mfdemo}). Again, all experimental curves show an good level of
agreement with the analytic predictions.

\subsection{Towards a realistic selection function}
\label{sec:rs}

The idealised situation analysed in the previous section, where astronomical
sources match exactly the template used for the filtering, is never encountered
in the reality of blind surveys.  Each real world emission line source can have
different spectral and spatial profiles.  A motivated strategy for the line
search is to opt for a template that maximises the SN for a large fraction of
the desired sources in the final sample.  This may be achieved iteratively by
varying the template parameters and checking the SN response of sources found in
a previous iteration or even a previous survey in the same field (see Fig.~7 of
\citetalias{Herenz2017}).  We will introduce a metric to quantify the loss in SN
with respect to the source matching filter below (Eq.~\ref{eq:45}).  If this
loss is undesirable high for a large fraction of the targeted population, one
might even consider to construct a catalogue from multiple filter templates.  In
this case, additional bookkeeping is required to track which template resulted
in a source entering the catalogue, and multiple entries for an individual
sources recovered with different templates need to be weeded out.  We will not
consider this case further here.  The aim of this section is rather to provide
ideas towards answering question: ``\textit{What is the realistic selection
  function for a given source population in the catalogue that results from the
  search with a template $\bm{S}$?}''.  We understand as ``source population''
here a class of emission line objects, e.g., high-redshift LAEs, whose spatial
and spectral properties are understood in a statistical sense.

\subsubsection{Formalism for source-template mismatches}
\label{sec:form-source-templ}

In order to find a realistic selection function as defined above we first have
to analyse what happens to $f_C$ for a single type of source
that is that is not matched by the search template.  To this aim we have to
calculate the response of the 3D filtering operation in Eq.~\eqref{eq:26} to the
non-template matching source.  We start by defining the voxels of the
non-template matching source profile, $\bm{\widetilde{S}}$, where the spatial
and spectral components are independent of each other
\begin{equation}
  \label{eq:37}
  \widetilde{S}_{x,y,z} = \widetilde{S}_{xy}^\mathrm{spat}
\widetilde{S}_{z}^\mathrm{spec} \;\text{,}
\end{equation}
where $\widetilde{S}_{ij}^\mathrm{spat} \neq {S}_{ij}^\mathrm{spat}$ and
$\widetilde{S}_{k}^\mathrm{spec} \neq {S}_{k}^\mathrm{spec}$ for
${S}_{ij}^\mathrm{spat}$ and ${S}_{k}^\mathrm{spec}$ from Eq.~\eqref{eq:16}.
Moreover, $\bm{S}$ is normalised:
$\sum_{ijk} \widetilde{S}_{ijk} \equiv \sum_{ij}
\widetilde{S}_{ij}^\mathrm{spat} \equiv \sum_k \widetilde{S}_{k}^\mathrm{spec}
\equiv 1$.  We now insert this source at position $x',y',z'$ into a data cuboid
that contains otherwise only noise.  In the spirit of Eq.~\eqref{eq:33} we write
\begin{equation}
  \label{eq:38}
  F_{x,y,z} = \xi \cdot A \cdot \widetilde{S}_{x-x',y-y'}^\mathrm{spat}
  \widetilde{S}_{z-z'}^\mathrm{spec} + n_{x,y,z} \; \text{.}
\end{equation}
with the amplitude $A$ as defined in Eq.~\eqref{eq:35}.  Here we introduced the
factor $\xi$ that will provide us with the required rescaling of the line flux,
$\xi \cdot F_\mathrm{line}$, in order to make the expectation value of the
filter in Eq.~\eqref{eq:26} at position $x',y',z'$ for the source given in
Eq.~\eqref{eq:38},
\begin{equation}
  \label{eq:39}
  E_\xi \left [\mathcal{T}^{(x',y',z')} \right] =
  \frac{\xi \cdot A^2}{E_\mathcal{I}\left[\mathcal{T}^{(x',y',z')}\right]} \cdot \sum_k
  \frac{S_k^{\{z'\}} \widetilde{S}_{k}^\mathrm{spec}}{\sigma^2_{z'-k}} \cdot
  \sum_{ij} S_{ij}^{\{z'\}} \widetilde{S}_{ij}^\mathrm{spat} \;\text{,}
\end{equation}
identical to the expectation value $E_\mathcal{I}[\mathcal{T}^{(x',y',z')}]$
from Eq.~\eqref{eq:34}.  Formally, we thus require
\begin{equation}
  \label{eq:40}
  E_\xi[\mathcal{T}^{(x',y',z')}] \stackrel{!}{=}  E_\mathcal{I}[\mathcal{T}^{(x',y',z')}] \;\text{,}
\end{equation}
and evaluating this condition results in
\begin{equation}
  \label{eq:41}
  \xi(\lambda[z]) = 
  \cfrac{ \sum_n \cfrac{ \left ( S_n^{\{z \}}  \right)^2  }{ \sigma_{z-n}^2}
    \cdot \sum_{ij} \left ( S_{ij} \right )^2 } 
   {\sum_n \cfrac{S_n^{\{z \}} \widetilde{S}^\mathrm{spec}_n}{\sigma_{z-n}^2}
    \cdot \sum_{ij} S_{ij} \widetilde{S}_{ij}^\mathrm{spat} } 
\end{equation}
as expression for the required flux rescaling factor.
Moreover,
denoting with $C_{\widetilde{S}}$ the proportionality factor in
units of inverse flux (Eq.~\ref{eq:27}) that needs to be used in the selection function
(Eq.~\ref{eq:31}) or for the 50\%-completeness limit (Eq.~\ref{eq:32}), and
noting that Eq.~\eqref{eq:41} can also be expressed as
\begin{equation}
  \label{eq:42}
  \xi =
  \frac{E_\xi^2}{A^2} \Bigg / \frac{E_\xi E_\mathcal{I}}{A^2} =
  \frac{E_\mathcal{I}}{E_\xi} \; \text{,}
\end{equation}
were we dropped the arguments of the expectation values and $\xi$, we also
recover
\begin{equation}
  \label{eq:43}
  \xi(\lambda) = \frac{C_\mathcal{I}(\lambda)}{C_{\widetilde{S}}(\lambda)} 
  \quad \text{or} \quad  C_{\widetilde{S}}(\lambda) = \frac{
    C_\mathcal{I}(\lambda)} {\xi(\lambda)} \; \text{.}
\end{equation}
The rescaling of the flux axis of the ideal selection function with $\xi$ from
Eq.~\eqref{eq:41} or the rescaling of the proportionality factor
$C_\mathcal{I}(\lambda)$ from Eq.~\eqref{eq:36} with $\xi^{-1}$
(Eq.~\ref{eq:43}) are of practical relevance for the calculation of a
non-template matching source's selection function.

It is, moreover, also of interest to quantify the relative loss in SN at a fixed
$F_\mathrm{line}$ (i.e., $\xi \equiv 1$) due to the source-template mismatch
with respect to the SN that would be obtained for a filter that is perfectly
matched to the source.  In analogy to Eq.~\eqref{eq:34} we can write the optimal
expectation value for the SN of the source in Eq.~\eqref{eq:38} as
\begin{equation}
  \label{eq:44}
  E_\mathcal{\widetilde{S}} \left [ \mathcal{T}^{(x',y',z')} \right ] =  A \cdot \sqrt{ \sum_{ij} \left ( \widetilde{S}_{ij}
    \right )^2 } \cdot \sqrt{\sum_{k}  \frac{\left ( \widetilde{S}_k  \right
      )^2}{\sigma_{z'-k}^2} } \; \text{.}
\end{equation}
Denoting with $E_{\xi = 1} \left [ \mathcal{T}^{(x',y',z')} \right ]$ the
expectation value from Eq.~\eqref{eq:39} for $\xi = 1$ the loss in SN relative
to the matched filter for $\bm{\widetilde{S}}$ is then given by the ratio
\begin{align}
 {}& \zeta(\lambda[z']) = \frac{E_{\xi = 1} \left [
    \mathcal{T}^{(x',y',z')} \right ] }{ E_\mathcal{\widetilde{S}}\left [
    \mathcal{T}^{(x',y',z')} \right ]} \label{eq:ins}  \\
 &= \cfrac{\sum_n \frac{S_n^{\{z' \}} \widetilde{S}^\mathrm{spec}_n}{\sigma_{z'-n}^2}
    \cdot \sum_{ij} S_{ij} \widetilde{S}_{ij}^\mathrm{spat} }{\sqrt{\sum_n
      \frac{ \left ( S_n^{\{z' \}}  \right)^2  }{ \sigma_{z'-n}^2}} \sqrt{\sum_{ij}
    \left ( S_{ij}^{\{z'\}} \right )^2 } \sqrt{\sum_{n}  \frac{\left ( \widetilde{S}_n  \right
      )^2}{\sigma_{z'-n}^2} } \sqrt{ \sum_{ij} \left ( \widetilde{S}_{ij}
    \right )^2 } } \;\text{.}   \label{eq:45}
\end{align}
This quantity provides a figure of merit to judge the quality of the applied
template for a particular source.

Equation~\eqref{eq:41} and Eq.~\eqref{eq:45} have to be evaluated numerically
for any source profile described by Eq.~\eqref{eq:37} and any template profile
that is described by Eq.~\eqref{eq:17}.

\subsubsection{Source-template mismatches in the spatial and spectral domain}
\label{sec:special-cases}

In principle, we are not restricted by the stated requirement that the spectral
and spatial component of the coveted source profile are independent of each
other (Eq.~\ref{eq:37}).  The expectation values $E_\xi$ in Eq.~\eqref{eq:39}
and $E_\mathcal{\widetilde{S}}$ in Eq.~\eqref{eq:44} could be calculated
numerically for an arbitrary 3D profile, $\bm{\mathcal{\widetilde{S}}}$, and
from those expectation values then $\xi$ (Eq.~\ref{eq:42}) and $\zeta$ follow
(Eq.~\ref{eq:ins}).  However, the separation allows us especially to analyse the
expected effects of source template mismatches in the spectral and spatial
domain separately.  This also allows us to develop intuition for the magnitude
of the flux rescaling of the selection function and the corresponding expected
loss in SN for realistic profiles (see also
Sect.~\ref{sec:parametric-examples}).
 
We first consider the case where only the spatial part of the source profile
differs from the template, but where the spectral part is a perfect match
($\widetilde{S}_{ij}^\mathrm{spat} \neq {S}_{ij}^\mathrm{spat}$, but
$\widetilde{S}_{k}^\mathrm{spec} \equiv {S}_{k}^\mathrm{spec}$).  Then
Eq.~\eqref{eq:41} reduces to
\begin{equation}
  \label{eq:46}
  \xi(\lambda[z]) = 
  \frac{\sum_{ij} \left ( S_{ij}^{\{z\}} \right )^2}{\sum_{ij} S_{ij}^{\{z\}} \widetilde{S}_{ij}^\mathrm{spat}} \; \text{,}
\end{equation}
and Eq.~\eqref{eq:45} becomes
\begin{equation}
  \label{eq:47}
  \zeta(\lambda[z]) =
  \frac{\sum_{ij} S_{ij}^{\{z \} } \widetilde{S}_{ij}^\mathrm{spat} }
       {\sqrt{\sum_{ij} \left ( S_{ij}^{\{z \} } \right )^2  }  \sqrt{\sum_{ij} \left ( \widetilde{S}_{ij}^\mathrm{spat}
           \right )^2 }} \; \text{.}
\end{equation}
As a direct consequence of the spatially invariant variance (Eq.~\ref{eq:22})
the pure spatial mismatch of template and source profile leads to rescaling of
the selection function (Eq.~\ref{eq:46}) or loss in SN (Eq.~\ref{eq:47}) that is
independent of the variance.  Moreover, Eq.~\eqref{eq:46} and Eq.~\eqref{eq:47}
also describe the rescaling of the limiting flux and the loss in S/N due to
template mismatch for the detection of background limited sources via matched
filtering in imaging data if the variance can be assumed to be constant for
every pixel.

We can go one step further and provide analytical expressions for
Eq.~\eqref{eq:46} and Eq.~\eqref{eq:47} for the case where both the source and
the template are 2D Gaussian profiles that differ in their dispersions.
Neglecting sampling effects, i.e. we assume that the pixel size is
significantly smaller than the dispersions of the profiles, allows us to replace
the summations by integrals.  Thus, writing $\sigma_\mathrm{T}$ for the 2D
template dispersion and $\sigma_\mathrm{S}$ the 2D source dispersion we find
\begin{align}
  \label{eq:48}
  \sum_{ij} \left ( S_{ij} \right )^2  &\simeq   \frac{1}{4 \pi^2
    \sigma_\mathrm{T}^4} \int_{-\infty}^{+\infty} \int_{-\infty}^{+\infty}
    \exp \left ( - \frac{x^2 +
      y^2}{\sigma_\mathrm{T}^2} \right ) \,\mathrm{d}x \mathrm{d}y \notag \\
 & = \frac{1}{4 \pi \sigma_\mathrm{T}^2}
\end{align}
and
\begin{multline}
  \label{eq:49}
  \sum_{ij} S_{ij} \widetilde{S}_{ij}^\mathrm{spat} \simeq \\ \simeq
  \frac{1}{4 \pi^2 \sigma_\mathrm{T}^2 \sigma_\mathrm{S}^2}
  \int_{-\infty}^{+\infty} \int_{-\infty}^{+\infty} \exp \left ( - \frac{1}{2}
    \frac{\sigma_\mathrm{T}^2 + \sigma_\mathrm{S}^2}{\sigma_\mathrm{S}^2
      \sigma_\mathrm{T}^2} ( x^2 + y^2 )
  \right )   \,\mathrm{d}x \mathrm{d}y \\
\quad\;\,  = \frac{1}{2 \pi (\sigma_\mathrm{S}^2 + \sigma_\mathrm{T}^2)} \;\text{.}
\hfill 
\end{multline}
Then, inserting Eq.~\eqref{eq:48} and Eq.~\eqref{eq:49} into Eq.~\eqref{eq:46},
and introducing the 2D Gaussian template mismatch factor,
\begin{equation}
  \label{eq:50}
  \chi_\mathrm{2D} = \frac{\sigma_\mathrm{S}}{\sigma_\mathrm{T}} \;\text{,}
\end{equation}
we obtain 
\begin{equation}
  \label{eq:51}
  \xi =  \frac{\sigma_\mathrm{S}^2 + \sigma_\mathrm{T}^2}{2\sigma_\mathrm{T}^2} =
  \frac{1 + \chi_\mathrm{2D}^2}{2} 
\end{equation}
as the required flux rescaling factor of the non-template matching 2D Gaussian
to reach the same SN as a template matching 2D Gaussian.   Moreover, using
the results from Eq.~\eqref{eq:48} and Eq.~\eqref{eq:49} in Eq.~\eqref{eq:47}
leads to
\begin{equation}
  \label{eq:52}
  \zeta = \frac{2 \sigma_\mathrm{T} \sigma_\mathrm{S}}{\sigma_\mathrm{T}^2 +
    \sigma_\mathrm{S}^2} = \frac{2 \chi_\mathrm{2D}}{1 + \chi_\mathrm{2D}^2} 
\end{equation}
as the loss in SN at fixed flux.  Equation~\eqref{eq:52} was derived in a
slightly different way by \cite{Zackay2017} in the context of source detection
in co-added images (their Appendix~D).

Considering the case where only the spectral part of the source profile differs
from that of the template and where the spatial part is perfectly matched
($\widetilde{S}_{k}^\mathrm{spec} \neq {S}_{k}^\mathrm{spec}$ and
$\widetilde{S}_{ij}^\mathrm{spat} \equiv {S}_{ij}^\mathrm{spat}$) we find from
Eq.~\eqref{eq:41}
\begin{equation}
  \label{eq:53}
  \xi(\lambda[z])
  =
\sum_n \cfrac{\left(s_n^{\{z\}} \right)^2}{\sigma_{z-n}^2} \Bigg /
    \sum_n
    \cfrac{S_n^{\{z\}} \widetilde{S}_n^\mathrm{spec}}
    {\sigma_{z-n}^2}
  \; \text{,}
\end{equation}
and Eq.~\eqref{eq:45} reduces to
\begin{equation}
  \label{eq:54}
  \zeta(\lambda[z]) = \cfrac{\sum_n \frac{S_n^{\{z \}}
      \widetilde{S}^\mathrm{spec}_n}{\sigma_{z'-n}^2}}{
    \sqrt{\sum_n
      \frac{ \left ( S_n^{\{z \}}  \right)^2  }{ \sigma_{z-n}^2}}
    \sqrt{\sum_{n}  \frac{\left ( \widetilde{S}_n  \right
      )^2}{\sigma_{z-n}^2} } } \;\text{.}
\end{equation}
The rescaling of the flux and loss in SN for purely spectral template mismatches
in Eq.~\eqref{eq:53} and Eq.~\eqref{eq:54}, respectively, depend on wavelength
due to the spectrally varying variances.

To compute reference values for $\xi$ and $\zeta$ for 1D source-template
mismatches we assume for a moment that the variances are not varying over the
extent of the filters and the source, since this allows for analytical estimate
for 1D Gaussians (Eq.~\ref{eq:69}) of different dispersions.  The calculations
are completely analogous to the the 2D case in Eq.~\eqref{eq:48} and
Eq.~\eqref{eq:49}. We find
\begin{equation}
  \label{eq:55}
  \xi_\mathrm{ref} = \sqrt{\frac{\sigma_\mathrm{T,1D}^2 + \sigma_\mathrm{S,1D}^2}{2
      \sigma_\mathrm{T,1D}^2}} = \sqrt{\frac{1 + \chi_\mathrm{1D}^2}{2}}
\end{equation}
and
\begin{equation}
  \label{eq:56}
  \zeta_\mathrm{ref} =  \sqrt{\frac{2 \sigma_\mathrm{T,1D}^2
      \sigma_\mathrm{S,2D}^2}{\sigma_\mathrm{T,1D}^2 + \sigma_\mathrm{S,1D}^2}}
  = \sqrt{\frac{2 \chi_\mathrm{1D}}{1 + \chi_\mathrm{1D}^2}}
\end{equation}
where $\sigma_\mathrm{T,1D}$ and $\sigma_\mathrm{S,1D}$ are the dispersions of
the template and source 1D Gaussian, respectively, and
\begin{equation}
  \label{eq:57}
  \chi_\mathrm{1D} = \frac{\sigma_\mathrm{S,1D}}{\sigma_\mathrm{T,1D}}
\end{equation}
is the 1D Gaussian template mismatch factor.  The subscript ``ref'' in
Eq.~\eqref{eq:55} and Eq.~\eqref{eq:56} explicitly indicates that these values
are reference values that are derived in the absence of spectrally varying
variances.

We see that the flux rescaling (Eq.~\ref{eq:55}) and SN loss factor
(Eq.~\ref{eq:56}) for the 1D Gaussian mismatch vary as a square root of the
expressions that were derived for the 2D Gaussian mismatch (Eq.~\ref{eq:51} and
Eq.~\ref{eq:52}).  Thus, in the absence of spectrally varying variances the
selection function appears more robust against source-template mismatches in the
spectral domain than against source-template mismatches in the spatial domain.
This argument was already stated in \citetalias{Herenz2017}, but without the
formal justification provided here.

\begin{figure}
  \centering
  \includegraphics[width=0.5\textwidth,trim=0 5 0 0, clip=true]{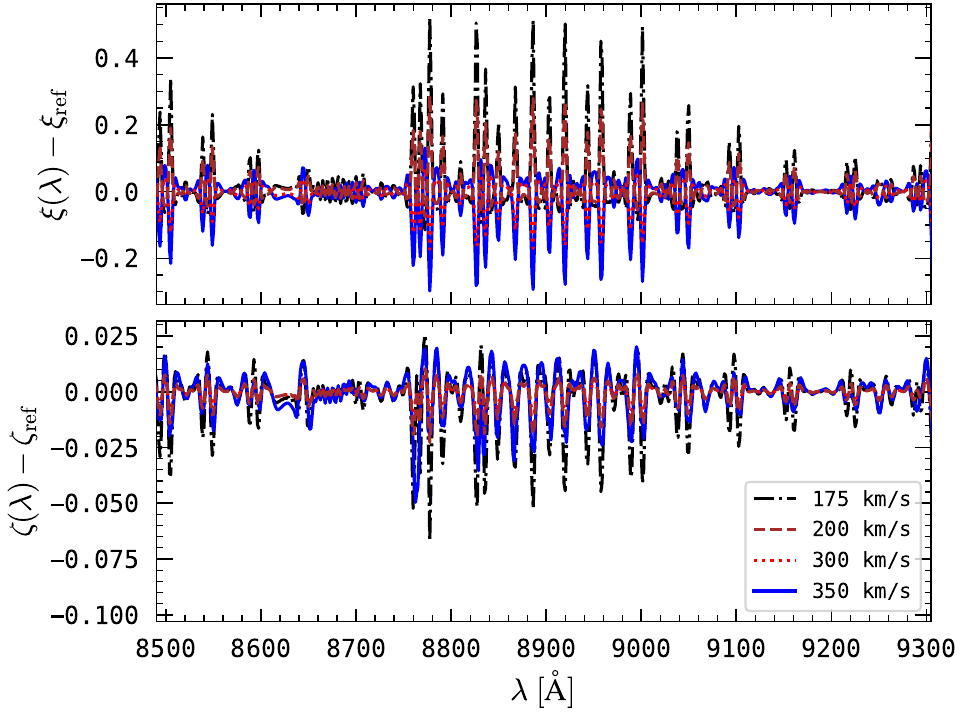}
  \caption{Illustration of the effects due to source-template mismatches in the
    spectral domain for 1D Gaussians (Eq.~\ref{eq:69}) that are not matched in
    width by the 1D Gaussian search template.  We show results for
    $v_\mathrm{FWHM}^\mathrm{template} = 250$\,km\,s$^{-1}$ and, as indicated in
    the legend,
    $v_\mathrm{FWHM}^\mathrm{source} = \{ 175, 200, 300, 350 \}$\,km\,s$^{-1}$.
    \textit{Top panel:} Difference between the flux rescaling factor,
    $\xi(\lambda)$ from Eq.~\eqref{eq:53}, and the reference rescaling factor,
    $\xi_\mathrm{ref}$ from Eq.~\eqref{eq:55}, that would be needed in the
    absence of spectrally varying variances. \textit{Bottom panel:} Difference
    between the loss in SN, $\zeta(\lambda)$ from Eq.~\eqref{eq:54}, and the
    reference value, $\zeta_\mathrm{ref}$ from Eq.~\eqref{eq:56}, that would
    follow in the absence of spectrally varying variances.  For the evaluation
    of Eq.~\eqref{eq:53} and Eq.~\eqref{eq:54} we here again used the effective
    variance spectrum of the MXDF (see Sect.~\ref{sec:verif-analyt-expr}), thus
    similar results are expected for the typical telluric background in the red
    part of the optical spectral range.}
  \label{fig:gm}
\end{figure}

Now it remains to be analysed how the mismatch between a 1D Gaussian filter and
source in the spectral domain are affected by a spectrally rapidly varying
background.  To illustrate this, we compare in Figure~\ref{fig:gm} the
difference between the calculated values for $\xi(\lambda)$ (Eq.~\ref{eq:53})
and $\zeta(\lambda)$ (Eq.~\ref{eq:54}) and the reference values
$\xi_\mathrm{ref}$ (Eq.~\ref{eq:55}) and $\zeta_\mathrm{ref}$ (Eq.~\ref{eq:56}).
We here analyse mismatches for
$v_\mathrm{FWHM}^\mathrm{template} = 250$\,km\,s$^{-1}$ and
$v_\mathrm{FWHM}^\mathrm{source} = \{ 175, 200, 300, 350 \}$\,km\,s$^{-1}$, that
is $\chi_\mathrm{1D} = \{ 0.7, 0.8, 1.2, 1.4\}$ and hence
$\xi_\mathrm{ref}= \{ 0.86, 0.91, 1.10, 1.21 \}$ and
$\zeta_\mathrm{ref}=\{ 0.969, 0.998, 0.992, 0.973\}$.  We show the results of
the calculations for the red part of the MUSE spectral range, where we have the
highest density and amplitudes of the sky emission lines.  Here we used again the
effective variance spectrum of the MXDF (Sect.~\ref{sec:verif-analyt-expr}).
This variance spectrum provides a good average of the sky-line amplitudes over
large window in time at Cerro Paranal, hence the results shown in
Figure~\ref{fig:gm} appear universal.

In the bottom panel of Figure~\ref{fig:gm} we find
$\zeta(\lambda) < \zeta_\mathrm{ref}$ in the vicinity of sky lines, i.e. at
fixed line flux the recovered SN is slightly lower than what is expected for a
constant background.  This is because the template does not correctly modulate
the shot-noise from the sky lines with respect to the source profile.  While the
difference $\zeta(\lambda) - \zeta_\mathrm{ref}$ is small, it is of comparable
amplitude as the reference value, $\zeta_\mathrm{ref}$, for the expected loss in
SN.  Of more practical relevance for the selection function is, nevertheless,
the difference $\xi(\lambda) - \xi_\mathrm{ref}$.  This difference is shown in
the top panel of Figure~\ref{fig:gm}.  We find $\xi(\lambda) > \xi_\mathrm{ref}$
for $v_\mathrm{FWHM}^\mathrm{source} < v_\mathrm{FWHM}^\mathrm{template}$ in the
vicinity of sky lines.  This is because the filter weighs the increased
background noise stronger in its wings compared to what would be appropriate
given the source profile.  The result is that a larger flux rescaling is
required with respect to the reference value at stationary noise.  On the other
hand, we have $\xi(\lambda) < \xi_\mathrm{ref}$ in the vicinity of sky-lines for
$v_\mathrm{FWHM}^\mathrm{source} > v_\mathrm{FWHM}^\mathrm{template}$.  This is
because the smaller width of filter template suppresses the variances in the
wings stronger than needed, while the flux from the wings of the source profile
does not contribute optimally to the signal.  Hence, slightly less flux
rescaling is needed in comparison to the reference value at stationary noise for
templates that are smaller then the source profile.

In summary, we find that in the presence of a rapidly varying spectral
background also spectral template mismatches need to be taken into account for a
correct estimate of the selection function.  Variations due to spatial template
mismatches will only dominate the required rescaling of the selection function
in the case of a slowly varying or constant background.

\subsubsection{Examples for realistic source-template mismatches in LAE searches}
\label{sec:parametric-examples}


We now analyse two more realistic source-template mismatches.  The examples
presented here appear of practical interest for emission line searches, and
especially searches for LAEs, in deep wide-field IFS datacubes.  As search
template we use here the template that was used for the search of faint
Ly$\alpha$ emitters in the MXDF (see \ref{sec:templ-param-used}; Herenz et al.,
in prep.).

\begin{figure}[t]
  \centering
  \includegraphics[width=0.45\textwidth,trim=0 10 0 5,clip=True]{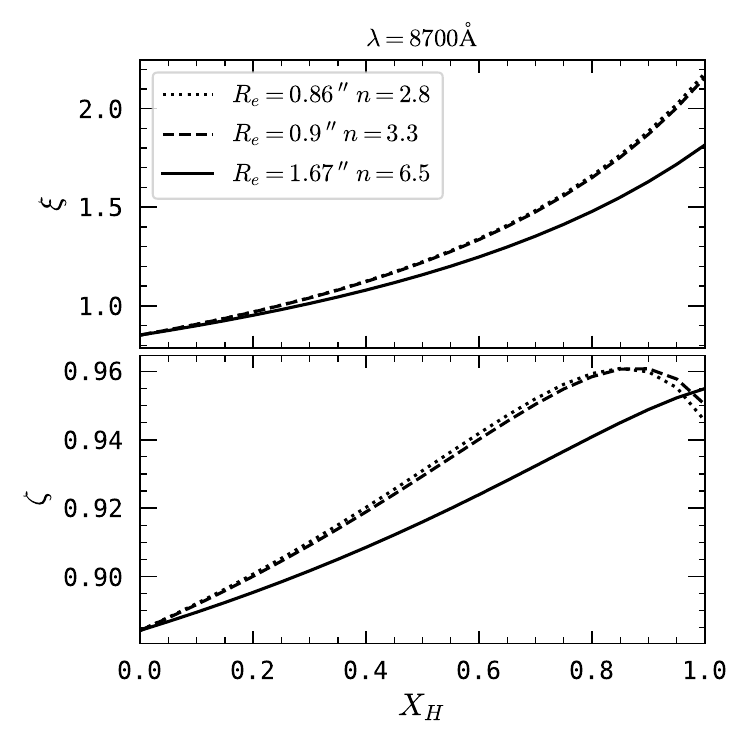}
  \caption{Rescaling ($\xi$, Eq.~\ref{eq:46}, \textit{top panel}) and loss in SN
    ($\zeta$, Eq.~\ref{eq:47}, \textit{bottom panel}) for average high-redshift
    LAE surface brightness profiles when recovered with the search template that
    was optimised for the recovery of known LAEs in the MXDF
    (Table~\ref{tab:sn2}).  Equation~\eqref{eq:46} and Eq.~\eqref{eq:47} were
    evaluated for this plot at $\lambda = 8700$\,\AA{}. The source surface
    brightness profiles are described PSF core + S\`{e}rsic halo according to
    Eq.~\eqref{eq:59}, where the parameters for the S\`{e}rsic halo
    (Eq.~\ref{eq:58}) are from \citeauthor{Wisotzki2018}
    (\citeyear{Wisotzki2018}; their Extended Data Table 1) as indicated in the
    legend.  The PSF component is modelled as a wavelength dependent Moffat
    (Eq.~\ref{eq:67}) according to Table~\ref{tab:mof}.  }
  \label{fig:xhserc}
\end{figure}

\paragraph{Example 1: Moffat core + S\'{e}rsic halo profiles}

The spatial distribution of Ly$\alpha$ emission from high-redshift LAEs can
almost always be characterised by a spatially unresolved ``point-like'' core
component and an extended halo component.  The physical nature of the extended
Ly$\alpha$ emission is not completely understood, but resonant scatterings of
Ly$\alpha$ photons in neutral halo gas are thought to be a dominant mechanism
\citep[e.g.,][]{Steidel2011,Hayes2013,Wisotzki2015,Mas-Ribas2016,Leclercq2017,LujanNiemeyer2022}.

We here follow \cite{Wisotzki2018} who parameterised the average halo profile of
these sources with a \cite{Sersic1968} profile\footnote{An English translation
  of the relevant chapter in the seminal \cite{Sersic1968} atlas, that was
  published in Spanish, can be found at
  \url{https://doi.org/10.5281/zenodo.2562394}.}:
\begin{equation}
  \label{eq:58}
  I_H(r) = I_H(0) \times \exp \left ( - \left [ \frac{r}{R_e} \right
    ]^{1/n} \right ) \; \text{.}
\end{equation}
Here the parameters $R_e$ and $n$ define the characteristic radius and the
kurtosis of the profile \citep{Graham2005}.  The observable profile of the
``point-like + halo'' model can then be written as
\begin{equation}
  \label{eq:59}
  \widetilde{S}(r,\lambda) = I_0 \cdot \mathrm{PSF}(r,\lambda) \star \left \{ (1 - X_\mathrm{H} ) \delta(r)
    + X_\mathrm{H} I_H(r) \right \} \; \text{,}
\end{equation}
where $\delta(r)$ is the 2D delta-function, $\star$ denotes the 2D convolution,
and $\mathrm{PSF}(r)$ denotes the model of the point spread
function\footnote{The effects of PSF convolution on the profile in
  Eq.~\eqref{eq:58} have been analysed by \cite{Trujillo2001a} and
  \cite{Trujillo2001} for a Gaussian and a Moffat PSF model, respectively.},
$I_0$ is a constant chosen such that the integral
$\int_0^{2\pi} \int_0^{\infty} S(r,\lambda) \, r^2 \,\mathrm{d}r\,\mathrm{d}\phi
\equiv 1$, and $X_\mathrm{H}$ ($\in [0,1]$) denotes the halo-flux fraction.
Here we model the PSF with a Moffat function (Eq.~\ref{eq:67} in \ref{sec:temp})
that was found to provide an adequate description of the point-spread function
(PSF) in adaptive-optics assisted MUSE observations \citep[][]{Fusco2020}.

We show the flux rescaling of the selection function, $\xi$ according to
Eq.~\eqref{eq:46}, and the loss in SN for this type of profile when recovered
with a 2D Gaussian (Table~\ref{tab:sn2}), $\zeta$ according to
Eq.~\eqref{eq:47}, as a function of the halo flux fraction $X_\mathrm{H}$ in the
top and bottom panel of Figure~\ref{fig:xhserc}, respectively.  The source
profiles are defined from the average LAE surface brightness profile parameters
derived by \cite{Wisotzki2018}.  In particular, we here evaluated
Eq.~\eqref{eq:46} at $\lambda = 8700$\,\AA{} for the profile given in
Eq.~\eqref{eq:59} with the Moffat PSF (Eq.~\ref{eq:67}) using the
parameterisation from \texttt{muse-psfr} in Table~\ref{tab:mof} and for three
different S\`{e}rsic profiles (Eq.~\ref{eq:58}) with
$R_e = (0.86\arcsec{}, 0.9\arcsec{}, 1.67\arcsec{})$ and $n=(2.8, 3.3, 6.5)$.
These parameters describe average LAEs in the redshift ranges $z = 3-4$,
$z=4-5$, and $z=5-6$, respectively \citep[][their Extended Data Table
1]{Wisotzki2018}.  We ensured that the numerical evaluations of
Eq.~\eqref{eq:58} and Eq.~\eqref{eq:59} were not affected by sampling
effects\footnote{The resulting profiles from those equations do not represent
  physical reality if the PSF and the S\'{e}rsic profile would only be evaluated
  at the centre of each spatial pixel, especially when $R_e$ is of similar
  dimensions as the spaxels and when $n$ becomes large \citep[see
  also][]{Peng2002}.  We note, that this is the case with the 2D S\'{e}rsic
  model provided in the Astropy package \citep{AstropyCollaboration2022} in the
  most recent version 5.1. The model
  \texttt{astropy.modeling.functional\_models.Sersic2D} must therefore not be
  used to model observed profiles.  Here we performed the calculations on a grid
  where 625 spatial pixels correspond to one MUSE spaxel, as we found that the
  results did not change significantly when using finer grids.  We also assumed
  a zero pixel-phase, i.e., the peak is located at the centre of the central
  pixel.  The so obtained 2D profiles of Eq.~\eqref{eq:59} were then downsampled
  to the native MUSE resolution, and those resampled 2D profiles were then used
  for the evaluation of Eq.~\eqref{eq:46}.}.

It can be seen in Figure~\ref{fig:xhserc} how PSF+halo profiles with halo flux
fraction of $\sim80\%-90\%$, which are typical for high-$z$ LAEs
\citep{Wisotzki2018}, are recovered at $\sim 140$\% of the flux of a perfect
template matching source.  Hence, the 50\% completeness limit for such sources
is slightly shallower than that of the idealised selection function.  Moreover,
the average $z>4$ LAEs surface-brightness profiles are recovered at 95\% of
their best possible SN ratio.  This demonstrates, that the 2D Gaussian profile
used for the LAE search in the MXDF (\ref{sec:templ-param-used}; Herenz et al.,
in prep.) is nearly optimal for recovering the average surface brightness
profiles of LAEs at the highest redshifts.

\paragraph{Example 2: Asymmetric Ly$\alpha$ line profiles}

The spectral profiles of Ly$\alpha$ emitting galaxies are of non-Gaussian
appearance due to resonant line scattering in their interstellar media
\citep[see lecture notes by][]{Dijkstra2017}.  A static scattering medium
results in a bi-modal wavelength distribution of photons that is symmetric
around the systemic emission.  An expanding medium, likely caused by galaxy
scale winds, leads to a more prominent red peak of this distribution and
absorption by the intergalactic medium can extinguish the blue peak of high-$z$
Ly$\alpha$ emitters \citep{Laursen2011}.

For a parametric model of this characteristic red-asymmetric ``saw-tooth''
spectral morphology we join two-half Gaussians with width parameters
$\sigma_\mathrm{b}$ and $\sigma_\mathrm{r}$ for the blue and red side,
respectively:
\begin{equation}
  \label{eq:60}
  s(\lambda) = \sqrt{\frac{2}{\pi}} \cdot \frac{1}{\sigma_\mathrm{b} + \sigma_\mathrm{r}} \times
  \begin{cases}
    \exp \left ( - \cfrac{(\lambda - \lambda_0)^2 }{2 \sigma_\mathrm{b}^2} \right )
    & \Leftrightarrow \lambda \leq \lambda_0 \\
    \exp \left ( - \cfrac{(\lambda - \lambda_0)^2 }{2 \sigma_\mathrm{r}^2} \right )
    & \Leftrightarrow \lambda > \lambda_0
  \end{cases} \;\text{.}
\end{equation}
Other impromptu parameterisations have been used in the literature\footnote{
  \cite{Mallery2012} and \cite{U2015} use a skewed normal distribution.
  However, \cite{Childs2018} found that the skewed normal does not allow for
  stringent constraints on the line profile morphology and especially the
  skewness in low-resolution ($R \lesssim 3000$) spectra at low SN.
  \cite{Shibuya2014} introduced an alternative functional form,
  $ s(\lambda) \propto \exp [ 1/2 \cdot (\lambda - \lambda_0)^2 / ( a(\lambda -
  \lambda_0) + d )^2 ]$, with the asymmetry parameter $a$ and the width
  parameter $d$.  However, the numerical instability at
  $\lambda = \lambda_0 - d/a$ and the dip of to zero around this location appear
  artificial.  Moreover, the range in $a$ that leads to realistic looking
  profiles with the \citeauthor{Shibuya2014} parameterisation is very
  narrow. For these reasons we here introduced a new function for parameterising
  Ly$\alpha$ profiles in Eq.~\eqref{eq:60}.}, but they do not posses special
qualities that would render them more useful for our purposes than
Eq.~\eqref{eq:60}.  As asymmetry parameter we use the blue-to-red flux ratio:
\begin{equation}
  \label{eq:61}
  a_\mathrm{f} = \frac{\int_{-\infty}^{\lambda_0} s(\lambda) \,
    \mathrm{d}\lambda}{\int_{\lambda_0}^{+\infty} s(\lambda) \,
    \mathrm{d}\lambda} \;\text{.}
\end{equation}
Such a working definition for characterising the Ly$\alpha$ line asymmetry was
also advocated by \cite{Rhoads2003} and \cite{Childs2018}.  Inserting
Eq.~\eqref{eq:60} into Eq.~\eqref{eq:61} we have
\begin{equation}
  \label{eq:72}
  a_\mathrm{f} = \frac{\sigma_\mathrm{b}}{\sigma_\mathrm{r}} \;\text{.}
\end{equation}
We furthermore
define the effective width of the asymmetric profile as
\begin{equation}
  \label{eq:62}
  \sigma_\mathrm{eff} = \frac{\sigma_\mathrm{b} + \sigma_\mathrm{r}}{2} \;\text{.}
\end{equation}
This definition implies
\begin{equation}
  \label{eq:63}
  \sigma_\mathrm{r} = \frac{2 \sigma_\mathrm{eff}}{a_\mathrm{f}+1} \quad \text{and} \quad
  \sigma_\mathrm{b} = \frac{2 a_\mathrm{f} \sigma_\mathrm{eff}}{a_\mathrm{f}+1}
  \; \text{.}
\end{equation}
For $a_\mathrm{f}=1$ the line-profile is a symmetric Gaussian, and for $a < 1$
we obtain an approximate line shape of the characteristic red-asymmetric line
profiles.  The observed line profile, $s_\mathrm{obs}$, results from a
convolution of $s(\lambda)$ in Eq.~\eqref{eq:60} with instruments line spread
function (LSF),
\begin{equation}
  \label{eq:64}
  s_\mathrm{obs}(\lambda) = I_0 \cdot \mathrm{LSF}(\lambda) * s(\lambda) \;\text{,}
\end{equation}
where $*$ denotes the 1D convolution and $I_0$ is chosen such that
$\int s_\mathrm{obs}(\lambda) \, \mathrm{d}\lambda \equiv 1$.  For simplicity,
we here adopt for the LSF a 1D Gaussian profile with dispersion
$\sigma_\mathrm{LSF}$, noting that the wings of the LSF in MUSE are more
pronounced than the wings of a Gaussian \citep[][their
Sect. 4.10]{Weilbacher2020}.  The width of the LSF over the wavelength range
varies smoothly from $v_\mathrm{FWHM}^\mathrm{LSF} \approx 190$\,km\,s$^{-1}$ in
the blue to $v_\mathrm{FWHM}^\mathrm{LSF} \approx 80$\,km\,s$^{-1}$ in the red.
This wavelength dependence of the LSF FWHM can be described adequately by a
quadratic polynomial and we use here the coefficients provided in Sect.~4.2.2 of
\cite{Bacon2022}.  By using $\sigma_\mathrm{eff}$ from Eq.~\eqref{eq:62} then the
width of the observed line profile from Eq.~\eqref{eq:64} can then be
approximated via
\begin{equation}
  \label{eq:65}
  \sigma_\mathrm{obs} \approx \sqrt{\sigma_\mathrm{eff}^2 +
    \sigma_\mathrm{LSF}^2} \;\text{.}
\end{equation}

\begin{figure}[t]
  \centering
  \includegraphics[width=0.45\textwidth,trim=0 10 0 0,clip=true]{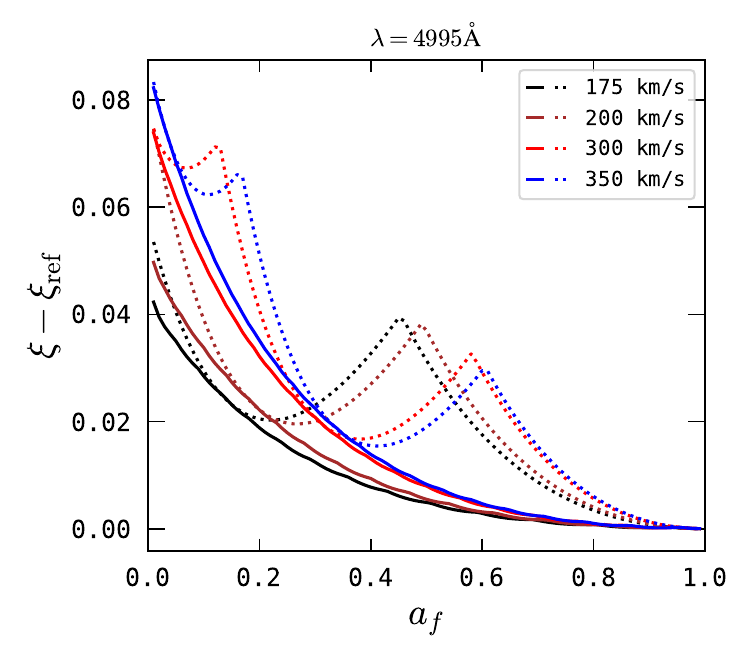}
  \caption{Effects of asymmetric Ly$\alpha$ profiles (Eq.~\ref{eq:60}) on the
    required flux rescaling of the selection function for searches with a
    symmetric 1D Gaussian profile.  We plot the difference
    $\xi - \xi_\mathrm{ref}$ as a function of the asymmetry parameter
    $a_\mathrm{f}$ (Eq.~\ref{eq:72}).  The solid lines are the evaluations of
    the relevant expressions on a grid that is 10-fold sampled with respect to
    the spectral sampling of MUSE, whereas the dotted lines show the results if
    the LSF convolved line profile (Eq.~\ref{eq:64}) is resampled to the MUSE
    grid (see text for details).  The different colours correspond to different
    intrinsic effective line widths (Eq.~\ref{eq:62}), with their
    $v_\mathrm{FWHM}$ provided in the legend.  }
  \label{fig:aslt}
\end{figure}

Above considerations now allow us to analyse the expected magnitude of the
required flux rescaling due to LSF-convolved asymmetric line profiles.  To put
this rescaling in context we compare it with reference rescaling for 1D
Gaussian-Gaussian template mismatches (Sect.~\ref{sec:special-cases}).  To
calculate this reference value, $\xi_\mathrm{ref}$, we use Eq.~\eqref{eq:55}
with $\chi_\mathrm{1D}$ from Eq.~\eqref{eq:57}, where we substitute
$\sigma_\mathrm{S,1D}$ with $\sigma_\mathrm{obs}$ from Eq.~\eqref{eq:65}.  To
compute the actual rescaling factor, $\xi$, we evaluate Eq.~\eqref{eq:53} with
the LSF convolved asymmetric profile according to Eq.~\eqref{eq:60} and
Eq.~\eqref{eq:64} for $0 < a_\mathrm{f} \leq 1$.  For the computation of
Eq.~\eqref{eq:65} we assume the noise to be constant over the width of the
filter and the source profile, as our aim here is to find the difference with
respect to the $\xi_\mathrm{ref}$, which was also computed under the same
assumption.  We show the results of this computation in Figure~\ref{fig:aslt}.
There we plot the difference $\xi - \xi_\mathrm{ref}$ as a function of the
asymmetry parameter $a_\mathrm{f}$.

Explicitly, the plot in Figure~\ref{fig:aslt} compares the magnitude of the flux
rescaling due to an asymmetric line profile with the rescaling due to a
Gaussian-Gaussian source-template mismatch if the width of the Gaussian source
corresponds to the effective observed width of the asymmetric profile.  We here
show the results for $\lambda = 4995$\,\AA{}, where
$v_\mathrm{FWHM}^\mathrm{LSF} \simeq 175$\,km\,s$^{-1}$, and we test asymmetric
line profiles of intrinsic effective widths (Eq.~\ref{eq:62})
$v_\mathrm{FWHM} = \{175,200,300,350\}$\,km\,s$^{-1}$ and thus, according to
Eq.~\eqref{eq:65},
$v_\mathrm{FWHM}^\mathrm{obs} \approx \{ 250, 265, 350, 390\} $\,km\,s$^{-1}$.
The width of the 1D Gaussian template profile is 250\,km\,s$^{-1}$
(\ref{sec:templ-param-used}).  We performed the computations first on a grid
that is 10-fold sampled with respect to the native MUSE resolution (i.e.,
$\Delta \lambda = 0.125$\,\AA{}; solid lines in Fig.~\ref{fig:aslt}), and we then
resampled the observed line profile to the native MUSE resolution
($\Delta \lambda = 1.25$\,\AA{}; dotted lines in Fig.~\ref{fig:aslt}).  The
oscillating behaviour of the resampled curves can be explained by a non-zero
pixel phase of the first-moment of the resampled line profile, since for
evaluation of Eq.~\eqref{eq:53} we always have to align the source profile with
the template profile such that the maxima of profile and template are at the
same pixel.  The pixel phase $p$, with $|p| \leq 0.5$, is defined as the
relative difference of the profiles true maximum with respect to the pixel
centre \citep[cf.][]{Robertson2017}.  The peaks in the curves for the resampled
profiles occur when the the pixel phase difference of the true peak with respect
to the peak pixel in the resampled profile are maximal.

We performed above analysis over the whole wavelength range of MUSE and we found
that the quantitative and qualitative behaviour of the curves shown in
Figure~\ref{fig:aslt} are not altered significantly.  Compared to
Gaussian-Gaussian template mismatches, the required additional flux rescaling
due to asymmetric lines is always below 10\,\%.  When analysing the loss of SN
according to Eq.~\eqref{eq:54} in comparison to the reference loss for 1D
Gaussian-Gaussian source-template mismatches (Eq.~\ref{eq:56}) we found that the
difference is always $\lesssim 1$\,\%.  We thus conclude, that the effect of
asymmetric line profiles is only relevant for profiles that are highly
asymmetric and whose observed line profiles are significantly broader than the
template profile.  Hence, the dominant effect for the rescaling of the selection
function for asymmetric line profiles is due to a mismatch of their line widths
with respect to the template profile and the rescaling formalism for 1D
Gaussian-Gaussian source-template mismatches developed in
Sect.~\ref{sec:special-cases}, which also incorporates the effect of spectrally
varying variances, encapsulates the required rescaling and loss in SN
satisfactorily.

\subsubsection{Synthesis for population studies}
\label{sec:synth-popul-stud}

Above considerations provide us with $C_{\widetilde{S}}(\lambda)$, i.e. the
$C$-factor for calculating the selection function (Eq.~\ref{eq:31}) or the 50\%
completeness limit (Eq.~\ref{eq:32}) for a non-template matching source
$\widetilde{\bm{S}}$ described by Eq.~\eqref{eq:37}.  We recall that
$C_{\widetilde{S}}(\lambda)$ can be expressed by a rescaling of $C$-factor
(Eq.~\ref{eq:43}) of the idealised selection function ($C_\mathcal{I}(\lambda)$;
Sect.~\ref{sec:ideal-select-funct}) with the rescaling factor $\xi(\lambda)$
according to Eq.~\eqref{eq:41}.  We remark, that the numerical evaluation of the
relevant expressions is nearly instant, even when considering the required
sub-sampling and re-binning schemes that are needed for dealing with S\'{e}rsic
type profiles.  For the calculation of a realistic selection function for a
population of sources it appears thus feasible to consider large ensembles of
source templates $\widetilde{S}_i$ from some population and then to calculate
$C_{\widetilde{S}_i}$ for each of those sources.  Then, Eq.~\eqref{eq:31} needs
to be evaluated for each $C_{\widetilde{S}_i}$ to obtain
$f_{C_{\widetilde{S}_i}}$, and the final selection function is then given by
$ f = \sum_i w_i f_{C_{\widetilde{S}_i}}$, where the weights
($\sum_i w_i \equiv 1$) have to be chosen according to the expected occurrence
rate of $\widetilde{S}_i$ within the population.  Alternatively, the source
parameters or profiles are directly drawn from the underlying population, and in
this case $f = N^{-1} \sum_i f_{C_{\widetilde{S}_i}}$, where $N$ is the number
of draws.  The resulting selection function then provides a robust estimate of
the realistic selection function for the catalogue entries of such a population
in a IFS data cuboid.


\section{Summary and Concluding Remarks}  
\label{sec:sc}

This article presented a discrete matched filtering approach in three dimensions
that correctly accounts for rapidly varying variances along one dimension while
in the two other dimensions the variances are fixed.  The motivation for this
approach is the search for faint astronomical emission line signals in wide
field integral field spectroscopic datasets.  An implementation of the method is
provided in an updated version in the open source Python software \mbox{LSDCat}
that can be obtained from the Astrophysics Source Code Library:
\url{https://ascl.net/1612.002}.  We demonstrated in this article, by making use
of the publicly available MUSE-Wide Data Release 1, that the updated algorithm
provides indeed better SN for emission line signals in the spectral vicinity of
telluric air-glow lines.

As with the original \mbox{LSDCat}, the matched filtering routines are
implemented in \texttt{lsd\_cc\_spatial.py} and \texttt{lsd\_cc\_spectral.py}.
For \texttt{lsd\_cc\_spatial.py}, which computes the inner sum of
Eq.~\eqref{eq:26}, only a flux data cuboid and the template parameters
(\ref{sec:temp}) are needed as input.  The outer sum of Eq.~\eqref{eq:26} is
implemented in \texttt{lsd\_cc\_spectral.py}, and here the resulting temporary
cuboid from the inner sum and the effective variance spectrum are required as an
input.  If desired, the user can still use the classic algorithm
(Eq.~\ref{eq:18}) in order to reproduce results obtained with the original
implementation of LSDCat.

We discussed a useful property of the emission line search with a
matched-filter, namely that in such a search the selection function is
deterministic for a given variance spectrum provided that the noise is behaving
according to the expectations.  Under this provision the selection function can
be expressed by a simple analytic formula (Eq.~\ref{eq:31}), with a factor of
proportionality, $C(\lambda)$, that only depends on the congruence between
assumed source profile and actual source profiles.  We provide an expression for
$C(\lambda)$ for the idealised case where template and source match exactly
(Eq.~\ref{eq:36}).

We also presented ideas regarding how the deterministic selection function for
the idealised case can be used to obtain realistic selection functions for line
source profiles that are not fully congruent with the template.  In particular,
we provided a formula that allows to calculate the effect on the factor
$C(\lambda)$ for source profiles that differ from the template profiles
(Eq.~\ref{eq:47}).  We then analysed three example situations of spatial profile
mismatches.  We put the idea forward that this method can be extended to obtain
realistic selection functions for catalogue entries of a population of emission
line sources, where the spatial and spectral properties have been understood in
a statistical sense.

Using an algorithm with a deterministic selection function removes the need for
computationally cumbersome source insertion and recovery experiments.  Moreover,
such an algorithm may also be used to efficiently design IFS survey
observations.  State-of-the art telescope facilities require a reliable estimate
of the required exposure time already at the application stage.  To this aim
interactive exposure time calculators are provided.  The calculations of those
tools use a model of the detectors and the atmosphere to provide a reliable
estimate of the background noise.  Such estimates of the background noise could
then be used with the here presented approach to predict, e.g., catalogue
incidences of particular emission line galaxies given their line luminosity
function.

Here we verified the analytic selection function with $C(\lambda)$ for the
idealised case against a source insertion experiment.  This experiment was
carried out in the data cuboid from the deepest MUSE survey ever obtained, the
$t_\mathrm{exp} = 141$\,h MUSE eXtreme Deep Field.  In this dataset the analytic
expression for an idealised selection function, where the sources match the
template exactly, shows excellent congruence with a selection function derived
from a source insertion and recovery experiment.  We remark that the MXDF
dataset is somewhat special in the sense that the observing strategy was chosen
to homogenise the background noise and to remove any potential residual
systematics.  Nevertheless, the assumption of spatially invariant noise is also
met to a sufficient degree in shallower MUSE data, and it was shown for the
standard dithering strategy of the deep MUSE observations of the MUSE Hubble
Deep Field South survey \citep{Bacon2015} that the noise scales only with some
moderate deviation from the expected $1/\sqrt{N_\mathrm{exp}}$ behaviour.
Nevertheless, a full verification of the here presented approach with shallower
data cuboids is desirable in the future.

The assumption of spatially invariant noise will not be met at positions where
bright source introduce shot-noise themselves.  Even if their flux could be
subtracted perfectly, e.g., with the methods presented by \cite{Kamann2013} for
stars and \cite{Schmidt2019} for galaxies, the spaxels covered by those sources
will violate the assumption of a background limited search.  Therefore the SN
values computed with the methods presented here will be biased low.  For
population studies, that rely on accurate selection functions it is therefore
necessary to identify such regions and to exclude them from the analysis.  In
MUSE data cuboids this can be achieved, e.g., by visual inspection of the
variance cube.  However, most of the regions on the sky were searches for faint
emission line galaxies are performed, are typically chosen such that the density
of bright foreground sources is low.

Some important aspects of the line source detection problem in integral field
spectroscopic data were not discussed here.  We mention those briefly below.

First, we did not make direct statements regarding the reliability (sometimes
also dubbed purity) on the samples from our catalogue.  The reliability $R$ is
defined as the complement of the false detection probability $p_\mathrm{F}$:
$R = 1 - p_\mathrm{F}$ \citep[e.g.][]{Hong2014}.  As noted in
Sect.~\ref{sec:mf}, the standard conversion between $p_\mathrm{F}$ and detection
threshold is formally not correct.  A series of articles addressed this issue in
2D \citep{Vio2016,Vio2017,Vio2019}.  While the details are mathematically
involved, the upshot is that the detection threshold has to be chosen more
conservatively than what the standard expectation based on a Gaussian
distribution would provide.  An often used empirical approach to quantify the
reliability is to use negated datasets, which works well under the assumption of
the noise being symmetric \citep[see also][]{Serra2012}.

Second, the construction of the selection function is based on the ground truth
for $F_\mathrm{line}$, which however is unknown for observed sources, where we
measure $F_\mathrm{line}^\mathrm{obs} \pm \Delta F_\mathrm{line}^\mathrm{obs}$
to characterise the distribution of possible line fluxes.  This needs to be
taken into account, when we use the selection function for modelling purposes,
and the well known Eddington-Malmquist bias in luminosity function
determinations results from this effect \citep[e.g., Chapter 5.5
in][]{Ivezic2014}.  One way to avoid such biases are additional cuts of the
sample by only using sources where the errors on the flux measurements do not
correspond to significant changes in the selection function, i.e., where
$f_C(F_\mathrm{line}^\mathrm{obs} \pm \Delta F_\mathrm{line}^\mathrm{obs},
\lambda) \simeq f_C(F_\mathrm{line}^\mathrm{obs}, \lambda)$.  However, this may
drastically reduce the sample, and an alternative is to account for the
measurement errors in the modelling process \citep[][]{Rix2021}.

Last, a current bottleneck in the construction of line emitter catalogues based
on wide-field IFS data is the manual classification step.  Related to this are
spurious line signals due to imperfect continuum subtraction of sources with
significant continuum emission.  These residuals also have to be weeded out
manually from catalogues.  We advocate future research on those latter issues by
investigating machine learning techniques for classification and better
continuum subtraction methods that do not leave strong residuals
\citep[see][]{Kamann2013,Schmidt2019}.

While matched filtering is a useful method to uncover the faintest emission line
galaxies in wide-field IFS datasets, ultimately it is only the first step in the
detection and analysis chain, and subsequent steps are required before
astrophysical facts can be inferred from the data.  Arguably, however, a good
understanding of this first step in the chain is required for successful
statistical analyses of IFS survey data.

\section*{Acknowledgements}

I acknowledge that this paper benefited greatly from the careful work of a
anonymous referee.  I wish to express my thankfulness to Lutz Wisotzki for
sparking off the initial idea for the work presented here.  I thank Leindert
Boogaard from the Max-Planck-Institut f\"{u}r Astronomie and Friedrich Anders
from the Departament de F\'{i}sica Qu\`{a}ntica i Astrof\'{i}sica at the
University of Barcelona for careful proofreading the manuscript prior to
submission.  My special thanks go to the staff working at ESO/Chile and the
Paranal Observatory, especially Fernando Selman and Fuyan Bian, for a
interesting and educational time.  I wish also to express my gratitude to
Pascale Hibon for organising the ``Joint Observatories Kavli Science Forum in
Chile'', where I could discuss the ideas expressed in this article publicly for
the first time \citep{2022joks.confE..12H}.  Lastly, I thank Roland Bacon for
envisioning and leading the MUSE consortium and I thank all members (too many to
be mentioned individually here) for insightful discussions.

\normalsize


\appendix

\section{Description of the templates used in LSDCat}
\label{sec:temp}

For completeness and in order to contextualise the template parameters
introduced in Sect.~\ref{sec:selfun} we provide here a short description of the
3D template \textbf{S} used in LSDCat.  A more comprehensive description is provided in
\citetalias{Herenz2017}.

LSDCat templates are optimised for spatially unresolved sources in ground based
observations.  The spatial profile of such an emission line source can either be
approximated by a circular symmetric Gaussian,
\begin{equation}
  \label{eq:66}
  S_{x,y}(\lambda) = \frac{1}{2 \pi \sigma_\mathrm{G}^2(\lambda)} 
  \exp 
  \left ( 
    - \frac{x^2 + y^2}{2 \sigma_\mathrm{G}^2(\lambda)}
  \right )
  \;\text{,}
\end{equation}
with the dispersion $\sigma_\mathrm{G}$, or by a \cite{Moffat1969} profile,
\begin{equation}
  \label{eq:67}
  S_{x,y}(\lambda) = \frac{\beta - 1}{\pi r_\mathrm{d}^2(\lambda)} 
  \left [
    1 + \frac{x^2 + y^2}{r_\mathrm{d}^2(\lambda)}
  \right ]^{-\beta}
  \;\text{,}
\end{equation}
with the width parameter $r_\mathrm{d}$ and the kurtosis parameter $\beta$.  We
remark that equation~\eqref{eq:67} in the limit $\beta \rightarrow \infty$ is
identical to Eq.~\eqref{eq:66} \citep{Trujillo2001}.

The dependence on wavelength $\lambda$ of $\sigma_\mathrm{G}$ in
Eq.~\eqref{eq:66} or $r_\mathrm{d}$ in Eq.~\eqref{eq:67} is required, since the
spatial resolution of ground based observations is wavelength dependent
\citep[see, e.g.,][]{Hickson2014}.  Here this wavelength dependence of the
point-spread function is empirically modelled by a polynomial,
\begin{equation}
  \label{eq:68}
  \mathrm{FWHM}(\lambda)[\text{\arcsec}] = \sum_{i=0}^{N_p} p_i (\lambda - \lambda_0)^i
  \;\text{,}
\end{equation}
and already a quadratic polynomial ($N_p = 2$) provides quite an accurate
description over the optical wavelength range.  The relation between the full
width at half maximum (FWHM in Eq.~\ref{eq:68}) and $\sigma_\mathrm{G}$ in
Eq.~\eqref{eq:66} is $\sigma_\mathrm{G} = \mathrm{FWHM} / (2\,\sqrt{2 \ln 2})$,
whereas for the width parameter $r_d$ in Eq.~\eqref{eq:67} we have
$r_\mathrm{d} = \mathrm{FWHM} / \left (2\,\sqrt{2^{1/\beta} - 1} \right )$.
Several empirical ways to determine optimal values for the $p_i$ in
Eq.~\eqref{eq:68} have been developed for MUSE observations \citep[see,
e.g.,][]{Herenz2017a,Bacon2017,Urrutia2018}.  Moreover, for MUSE observations
with laser-assisted ground layer adaptive optics the wavelength dependence of
the Moffat function can be modelled from data provided by the adaptive optics
telemetry system \citep{Oberti2018} with the \texttt{muse-psfr} software
\citep{Fusco2020}.

We point out that the here adopted convention for the polynomial description of
the $\lambda$-dependence Eq.~\eqref{eq:68} differs from the convention in the
MUSE Python data analysis framework and in \texttt{muse-psfr}.  The recipe for
conversion is provided in \ref{sec:mpdaflsd}.  Lastly, we note that the parameter
$\beta$ in Eq.~\eqref{eq:67} is usually not strongly dependent on wavelength,
but in \mbox{LSDCat} also a polynomial analogous to Eq.~\eqref{eq:68} can be
used to enforce a wavelength dependence on $\beta$ if desired.

The spectral profile is modelled as a simple 1D Gaussian
\begin{equation}
  \label{eq:69}
  S_z = 
  \frac{1}{\sqrt{2\pi}\sigma_z}
  \exp \left 
    ( - \frac{z^2}{2\sigma_z^2}
  \right ) \; \text{,}
\end{equation}
whose width $\sigma_v$ (in km\,s$^{-1}$)is fixed in velocity space. For the
linear sampled wavelength grid,
\begin{equation}
  \label{eq:70}
  \lambda = z \cdot \Delta \lambda + \lambda_{z=0} \;\text{,}
\end{equation}
thus
\begin{equation}
  \label{lsd_eq:11}
  \sigma_z
  = 
  \frac{\sigma_v}{c}
  \left (
    \frac{\lambda_{z=0}}{\Delta \lambda} + z 
  \right ) \; \text{.}
\end{equation}
Here $\lambda_{z=0}$ denotes the wavelength at the first spectral bin ($z=0$) and
$\Delta \lambda$ denotes the wavelength increment per spectral bin ($\Delta
\lambda = 1.25$\AA{} is the default spectral increment for MUSE).

\section{Conversion of polynomial coefficients used in MPDAF and muse-psfr for use
  in LSDCat}
\label{sec:mpdaflsd}

The MUSE Python analysis framework \citep[MPDAF;][]{Bacon2016,Piqueras2017} and
\texttt{muse-psfr} \citep{Fusco2020} can model the wavelength dependence of the
width parameter of the Moffat function (Eq.~\ref{eq:67}).  However, the empirical
polynomial model adopted by those tools is written as
\begin{equation}
  \label{eq:a1}
  q(\lambda) = \sum_{i=0}^{N_p} b_i \left ( \frac{\lambda - \lambda_1}{\lambda_2
      - \lambda_1} - \frac{1}{2} \right )\; \text{,}
\end{equation}
whereas in LSDCat we adopt
\begin{equation}
  \label{eq:a2}
   p(\lambda) = \sum_{i=0}^{N_p} a_i (\lambda - \lambda_0)^i
\end{equation}
(cf.~Eq.~\ref{eq:68}).  Clearly, Eq.~\eqref{eq:a2} provides a more intuitive
description of the wavelength dependence, since $a_0$ refers to the parameter
under consideration (FWHM or $\beta$) at $\lambda_0$ (in \AA{}).  On the other
hand, Eq.~\eqref{eq:a1} appears numerically more accurate since $q$ evaluates on
the interval $[-1,1]$ where the density of digitally stored floating point
numbers is highest.  However, this extra level of numerical accuracy appears not
to provide practical benefits, at least for the use in LSDCat.  Using the
Heaviside step function,
\begin{equation}
  \label{eq:71}
  \Theta(k) =
  \begin{cases}
    1 & (k \leq 0) \\
    0 & (k > 0)
  \end{cases}
  \; \text{,}
\end{equation}
we can achieve
the conversion of the $b_i$ in Eq.~\eqref{eq:a1} to the $a_i$ in
Eq.~\eqref{eq:a2}  via
\begin{equation}
  \label{eq:a3}
  a_i = \sum_{j=0}^{N_p} c_j \, \binom{j}{i} \, \lambda_0^{i-j} \,  \Theta(j-i)
\end{equation}
with
\begin{equation}
  \label{eq:a4}
  c_i = \sum_{j=0}^{N_p} b_j \, \binom{j}{i} \alpha^i \beta^{i-j} \, \Theta(j-i) \;\text{,}
\end{equation}
where $\alpha$ and $\beta$ in Eq.~\eqref{eq:a4} are shorthands for
\begin{equation}
  \label{eq:a5}
  \alpha = \frac{1}{\lambda_2 - \lambda_1} \quad \text{and} \quad   \beta = - \left ( \frac{\lambda_1}{\lambda_2 - \lambda_1} + \frac{1}{2} \right
  )\,\text{.}
\end{equation}

\section{Template parameters used in the numerical examples}
\label{sec:templ-param-used}

\begin{table}
  \caption{\: Parameters describing the 3D Gaussian template used in for the LAE
    search MXDF (see \ref{sec:temp} for the parameterisation).  For the spatial
    domain the FWHM-$\lambda$ dependence is given as polynomaial coefficients of
    Eq.~\eqref{eq:68} with $\lambda_0 = 7050$\AA{}.}
  \label{tab:sn2}
  \centering
  \begin{tabular}{lc}
    \hline \hline
     \noalign{\smallskip}
    \multicolumn{2}{c}{\underline{spatial domain}} \\
    \noalign{\smallskip}
    \noalign{\smallskip}
    $p_i$ & FWHM [\arcsec \AA{}$^{-i}$] \\ 
    \noalign{\smallskip} \hline \noalign{\smallskip}
    $p_1$ & 1.0042 \\
    $p_2$ & $-3.322\times10^{-5}$ \\ 
    \noalign{\smallskip} \hline
    \noalign{\smallskip}
    \multicolumn{2}{c}{\underline{spectral domain}} \\
    \noalign{\smallskip}
    \noalign{\smallskip}
    $v_\mathrm{FWHM}$ & 250\,km\,s$^{-1}$ \\
    \noalign{\smallskip}
    \hline \hline
  \end{tabular}
\end{table}
\begin{table}
  \caption{~Polynomial coefficients in Eq.~\eqref{eq:a2} for
    $\lambda_0=7000$\,\AA{} that describe the wavelength dependence of a Moffat
    point-spread function model in the MXDF \citep{Bacon2022} according to
    \texttt{muse-psfr} \citep{Fusco2020}.}
  \label{tab:mof}
  \centering
  \begin{tabular}{lcc}
    \hline \hline \noalign{\smallskip}
    $p_i$ & FWHM [\arcsec \AA{}$^{-i}$] & $\beta$ \\
    \noalign{\smallskip}
    \hline
    $p_0$ & 0.49 & 1.96 \\
    $p_1$ & $-5.69 \times 10^{-5}$  & $-1.02\times10^{-4}$ \\ 
    $p_2$ & $\phantom{-}1.14 \times 10^{-9}$   & $\phantom{-}6.52\times10^{-10}$ \\
    $p_3$ & $\phantom{-}1.06 \times 10^{-12}$  & $\phantom{-}5.43\times10^{-12}$ \\
    $p_4$ & $-8.86 \times 10^{-17}$ & $-1.54\times10^{-15}$\\
    $p_5$ & $\phantom{-}2.91 \times 10^{-19}$  & $-3.75\times10^{-19}$\\
    \hline \hline
  \end{tabular}
\end{table}

For reference we list in Table~\ref{tab:sn2} and Table~\ref{tab:mof} the
template parameters that have been used for the calculations in
Sect.~\ref{sec:selfun}.  The templates and their parameterisations are listed
in~\ref{sec:temp}.

Table~\ref{tab:sn2} lists the parameters of the 3D Gaussian template that have
been used for a search of Ly$\alpha$ emitting galaxies (Herenz et al., in prep.)
in the recently released data cuboid of the MXDF (Bacon et al. 2022).  These
parameters were chosen to maximise the SN of known Ly$\alpha$ emitters from the
previous MUSE Ultra Deep Field survey \citep{Inami2017} where the footprint of
the MXDF is located.  As detection threshold we used
$SN_\mathrm{thresh} = 6.41$.  This template was used in the example of
Sect.~\ref{sec:verif-analyt-expr}, where we compute the idealised selection
function at $\lambda > 8500$\,\AA{} with Eq.~\eqref{eq:31} and
$C_\mathcal{I}(\lambda)$ from Eq.~\eqref{eq:36} and where we contrasted the
analytical calculation to a source insertion and recovery experiment.

Table~\ref{tab:mof} lists the polynomial coefficients that describe the
wavelength dependence of the Moffat \texttt{muse-psfr} \citep{Fusco2020} point
spread function model in the MXDF according to Eq.~\eqref{eq:a2} with
$\lambda_0 = 7000$\,\AA{}. The \texttt{muse-psfr} tool uses data that was
recorded by the adaptive optics telemetry system \citep{Oberti2018} during the
observations.  The polynomial coefficients are given in the header of the FITS
file of the MXDF datacube, but for the expression in Eq.~\eqref{eq:a1}.  We here
use the formalism provided in \ref{sec:mpdaflsd} to convert them into the more
natural form of Eq.~\eqref{eq:a2}.

\bibliography{lsd20_paper_revision_v2.bib}

\section*{Author Biography}

%
\begin{biography}{\includegraphics[width=60pt,height=70pt]{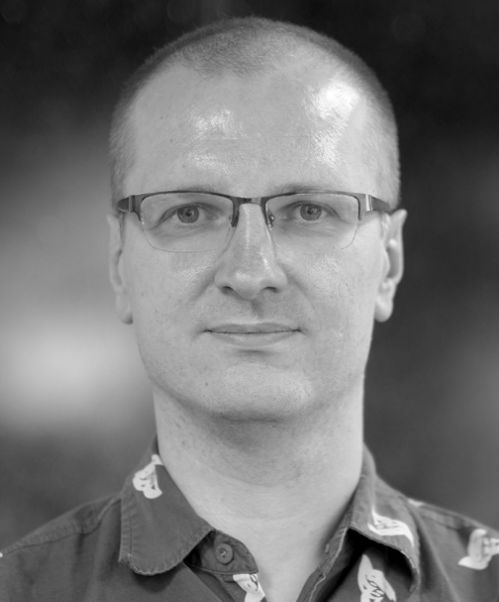}}{\textbf{Edmund
      Christian Herenz} is interested in extragalactic astrophysics and
    cosmology.  He obtained his PhD in 2016 from the University Potsdam for his
    thesis ``Detecting and understanding extragalactic Lyman $\alpha$ emission
    using 3D spectroscopy'' that was supervised by Lutz Wisotzki and Martin Roth
    at the Leibniz-Institute for Astrophysics Potsdam (AIP).  Afterwards he did
    a Post-Doc with Matthew Hayes at Stockholm University.  He was then awarded with
    a Fellowship from the European Southern Observatory where he supported
    operations of HAWK-I and MUSE at UT4 ``Yepun''.  He currently spends the 4th
    year of ESO Chile fellowship as a visiting researcher at the Leiden
    Observatory (Leiden University) in the Netherlands.}
\end{biography}

\end{document}